\documentclass[fleqn,usenatbib]{mnras}

\usepackage{newtxtext,newtxmath}

\usepackage[utf8]{inputenc}
\usepackage[T1]{fontenc}
\usepackage{ae,aecompl}

\usepackage{xcolor}

\usepackage{graphicx}	
\usepackage{amsmath}	
\usepackage{amssymb}	

\usepackage{siunitx}    
\usepackage{subfigure}  %
\usepackage{multirow}

\title[ARCHI - pipeline to study CHEOPS background stars]{ARCHI:  pipeline for light curve extraction of CHEOPS background stars}

\author[André M. Silva et al.]{
    \href{https://orcid.org/0000-0003-4920-738X}{\includegraphics[scale=0.75]{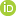}}André M. Silva$^{1,2}$\thanks{E-mail: amiguel@astro.up.pt},
    \href{https://orcid.org/0000-0001-9047-2965}{\includegraphics[scale=0.75]{Figures/ORCID-iD_icon-16x16.png}}Sérgio G. Sousa$^{1},$
    \href{https://orcid.org/0000-0003-4422-2919}{\includegraphics[scale=0.75]{Figures/ORCID-iD_icon-16x16.png}} Nuno Santos$^{1,2},$
    \href{https://orcid.org/0000-0001-7918-0355}{\includegraphics[scale=0.75]{Figures/ORCID-iD_icon-16x16.png}} Olivier D. S. Demangeon$^{1}$,
    \newauthor
    Pedro Silva$^{1,2}$, 
    S.Hoyer $^{3},$
    P.Guterman $^{3,4},$
    Magali Deleuil$^{3},$ 
    David Ehrenreich $^{5}$,
\\
$^{1}$Instituto de Astrofísica e Ciências do Espaço, Universidade do Porto, CAUP, Rua das Estrelas, 4150-762 Porto, Portugal \\
$^{2}$Departamento de Física e Astronomia, Faculdade de Ciências, Universidade do Porto, Rua Campo Alegre, 4169-007 Porto, Portugal\\
$^{3}$Aix Marseille Univ, CNRS, LAM, Laboratoire d’Astrophysique de Marseille, Marseille, France  \\
$^{4}$Division Technique INSU, BP 330, 83507 La Seyne cedex, France  \\
$^{5}$Observatoire astronomique de l’Université de Genève, 51 chemin des Maillettes 1290 Versoix, Switzerland \\
}
   
\date{Accepted XXX. Received YYY; in original form ZZZ}

\pubyear{2019}

\begin{document}
\label{firstpage}
\pagerange{\pageref{firstpage}--\pageref{lastpage}}
\maketitle

\begin{abstract}

High precision time series photometry from space is being used for a number of scientific cases. In this context, the recently launched \textit{CHEOPS} (ESA) mission promises to bring 20 ppm precision over an exposure time of 6 hours, when targeting nearby bright stars, having in mind the detailed characterization of exoplanetary systems through transit measurements. However, the official \textit{CHEOPS} (ESA) mission pipeline only provides photometry for the main target (the central star in the field). In order to explore the potential of CHEOPS photometry for all stars in the field, in this paper we present \textit{archi}, an additional  open-source pipeline module\footnotemark  to analyse the background stars present in the image. As \textit{archi} uses the official Data Reduction Pipeline
data as input, it is not meant to be used as independent tool to process raw CHEOPS data but, instead, to be used as an add-on to the official pipeline.
 We test \textit{archi} using \textit{CHEOPS} simulated images, and show that photometry of background stars in \textit{CHEOPS} images is only slightly degraded (by a factor of 2 to 3) with respect to the main target. This opens a potential for the use of CHEOPS to produce photometric time series of several close-by targets at once, as well as to use different stars in the image to calibrate systematic errors. We also show one clear scientific application where the study of the companion light curve can be important for the understanding of the contamination on the main target. 

\end{abstract}



\footnotetext{\url{https://github.com/Kamuish/archi}}
\section{Introduction}
    
    The CHaracterizing ExOPlanet Satellite, \textit{CHEOPS} \citep{Broeg2013,Fortier2014,Rando2018}, the recently launched ESA's mission is the first dedicated mission to better characterize planetary transits  with ultra-high precision photometry on stars known to have exoplanets.

    The orbit during the mission operation is a circular Sun-synchronous orbit, at an altitude of 800 km and an orbital period of approximately 100 minutes. The spacecraft is nadir locked and thus it will always be rotating around Earth, pointing towards the targeted direction. It has a 1024x1024 pixel Charge Coupled Device (CCD) but only a smaller window with a default size of 200x200 pixel is sent back to Earth. This region, or SubArray, is free to be centered anywhere on the CCD, allowing the possibility to select better regions of the full ccd for more precise photometry. The orbital configuration of the spacecraft  results in the rotation of the field of view and, consequently, the background stars rotate over the  CCD, at a rate of 3.6 deg/min. The satellite exhibits a small field of view, with a diameter of \ang{0.32}, a pixel scale of \textasciitilde 1'' and a Point Spread Function (PSF) with a radius of 12 pixels \cite{off_cheops_pipe}. It is expected that the targets have magnitudes in the 6 to 12 V-mag range, with an exposure time that can changed between 1 ms to 60s, depending on the target's brightness.

    This rotational movement will lead to the background stars occupying different pixels and, due to this constant change, we expect to find lower photometric precision when compared to the target star, that is maintained centered in a specific region of the CCD. On top of this, the exposure time is set taking into account the brightness of the central star, leaving the background ones in sub-optimal conditions, i.e. under or over exposed, either not being able to be detected or are saturated. Nevertheless there will be cases where the background stars observations can be still used to extract precise light curves, as we shall see later in this paper.

    In line with the main goals of the CHEOPS mission, the official Data Reduction Pipeline \citep{off_cheops_pipe},
    henceforth DRP, only provides photometry for the main target, i.e.
    the central star in the field. However in the same field of view we can
    find other stars, for which it may be scientifically useful to derive (for free)
    precise photometry timeseries. Precise photometry of the background
    stars can be used to diagnose systematic errors in the data and possibly allow to correct for spurious signals in the photometric timeseries of the main target, such as contamination from background stars, as shown in Section \ref{APP:contam_targ_bg}. If sufficient precision is achieved, this can in principle be used to confirm the astrophysical origin for observed signals. Examples include the contamination by background bright eclipsing binaries \citep{Abdul_Masih_2016,Deleuil_2018} and to diagnose other unexpected effects, including instrumentation-related. Finally, since 20\% of the available observing time of CHEOPS is open to the community (Guest Observing), other science cases, not related with exoplanet science, may gain from the observation of several stars in "dense" fields. Examples include photometry of open cluster stars that can't at present be analyzed using the official DRP. It is thus important to access the photometric precision that can be achieved for background stars in the CHEOPS field. This is the main goal of the present paper.

    We will start  by describing the DRP in Section \ref{Sec:DRP}. Afterwards, in Section \ref{Sec:methods} we shall present the algorithms that were implemented in \textit{archi}, our expansion to the DRP. Lastly, in Section \ref{sec:results} we shall see how \textit{archi} behaves under different observational conditions and benchmark the different routines. 

\section{The official Data Reduction Pipeline (DRP)} \label{Sec:DRP}

    Taking into account that this pipeline was built as an extension for the \textit{CHEOPS} mission DRP, 
    we must understand what this can do. As the pipeline is introduced in \cite{off_cheops_pipe}, we shall only give a brief introduction to the parts relevant to our work. 

    In a generalized way, we can describe the DRP as a collection of 3 different modules, that are applied in a sequential order, as presented below:

        \begin{enumerate}
            \item \textit{Calibration}: Corrects the instrumental response, by attempting to remove:
                \begin{itemize}
                    \item Bias and readout noise;
                    \item Non-linearity of the CCD;
                    \item Gain;
                    \item Dark current;
                    \item Flat Field.
                \end{itemize}
            \item \textit{Correction}: Corrects environmental effects, such as:
                \begin{itemize}
                    \item Smear correction;
                    \item Detection of bad pixels;
                    \item Detection and correction of cosmic rays;
                    \item Background.
                \end{itemize}
            \item \textit{Photometry}: Extracts the targeted star's light curve from the images.
        \end{enumerate}

        The background estimation within the DRP (pipeline) is performed with an histogram based method, that uses an anulus centered on the target, not counting the pixels that are in the aperture. The background value is then estimated by fitting a fitted skewed Gaussian and, afterwards, used to correct the full images. For further detail we refer to Section 5.3 of \cite{off_cheops_pipe}. 
        We can now look into the \textit{Photometry} module with greater detail, to understand the methodologies in use. The flux from the target star is calculated with a circular mask and, to maintain the same number of pixels in the mask, it is created once and afterwards it is shifted with an anti aliasing shifting algorithm, described below, to avoid altering the mask's surface. This pipeline, extracts the photometry of the target star with four different apertures:
        \begin{itemize}
            \item \textit{DEFAULT}: Uses a circular mask with a default radius of 33 pixels;
            \item \textit{OPTIMAL}: The mask aperture is optimized through the maximization of the signal to noise ratio;
            \item \textit{RINF}: The mask has a radius equal to 80 \% of the default mask size;
            \item \textit{RSUP}: The mask has a radius equal to 120 \%  of the default mask size.
        \end{itemize}

        To always have the mask centered over the central star, an iterative Gaussian apodization method is used to estimate the star's position in each image.
        This algorithm begins by placing a mask on the image position estimated by on-board software and removing from the images the contributions from nearby stars and noise caused by the jitter. Afterwards, the centre of light, from this corrected image, is calculated, and a new mask is centered in it. This process then repeats until a convergence criterium is met, which tends to occurs fairly quick, under 20 iterations. With this, it is capable of estimating positions with errors as low as $2x10^{-3}$ pix, as reported in \citep{off_cheops_pipe}. Due to those low errors we shall use those positions as one of \textit{archi}'s star tracking methods, as we shall see in Section \ref{Sec:star_track}.

    Lastly, since the mission is yet to see its first light, all obtained data sets are simulated, using CHEOPSim \citep{Art:cheops_Sim}, the official simulation tool.
    
\section{Methods} \label{Sec:methods}

\subsection{Finding the stars}
    Within this work we have developed two different methods to identify a star in the images. One of them makes use of the sky coordinates of the stars, the plate scale of the CCD and the rotation angle of the satellite to estimate positions within the image, Section ~\ref{Sec:init_fits} . The other one applies image processing techniques to extract the star's contour and it's location in the image, Section~\ref{Sec:dynam_init_track}. The former shall be referred to as the ``fits'' initial detection method, whilst the later is the ``dynam'' initial detection method.

    \subsubsection{Usage of a Star Catalogue} \label{Sec:init_fits}

        For \textit{CHEOPS} observations we need to define a star catalogue, available with the DRP processed data, from which we can extract the right ascension (RA) and declination (DEC) of each star. Furthermore, from the DRP's outputs, we know the scale of the images and the rotation angle of the satellite for each image. If we calculate the difference of RA and DEC between the target star and each non-target star and convert it to pixels, we know how far away each star is from the center. 
        
        Now that we know the separation, in pixels, we still need to take into account one last detail:  the RA and DEC were calculated for a CCD with a roll angle of zero, which is not guaranteed to happen on the first image of the data   set. Thus, we have to rotate the estimated points by an angle of $360 - \theta_{initial}$ to place them at the correct locations in the initial image, whit $\theta initial$  being defined as the rotation angle of the satellite for the first image in the Data Set.

            \begin{figure}
                    \centering
                \includegraphics[width=8cm]{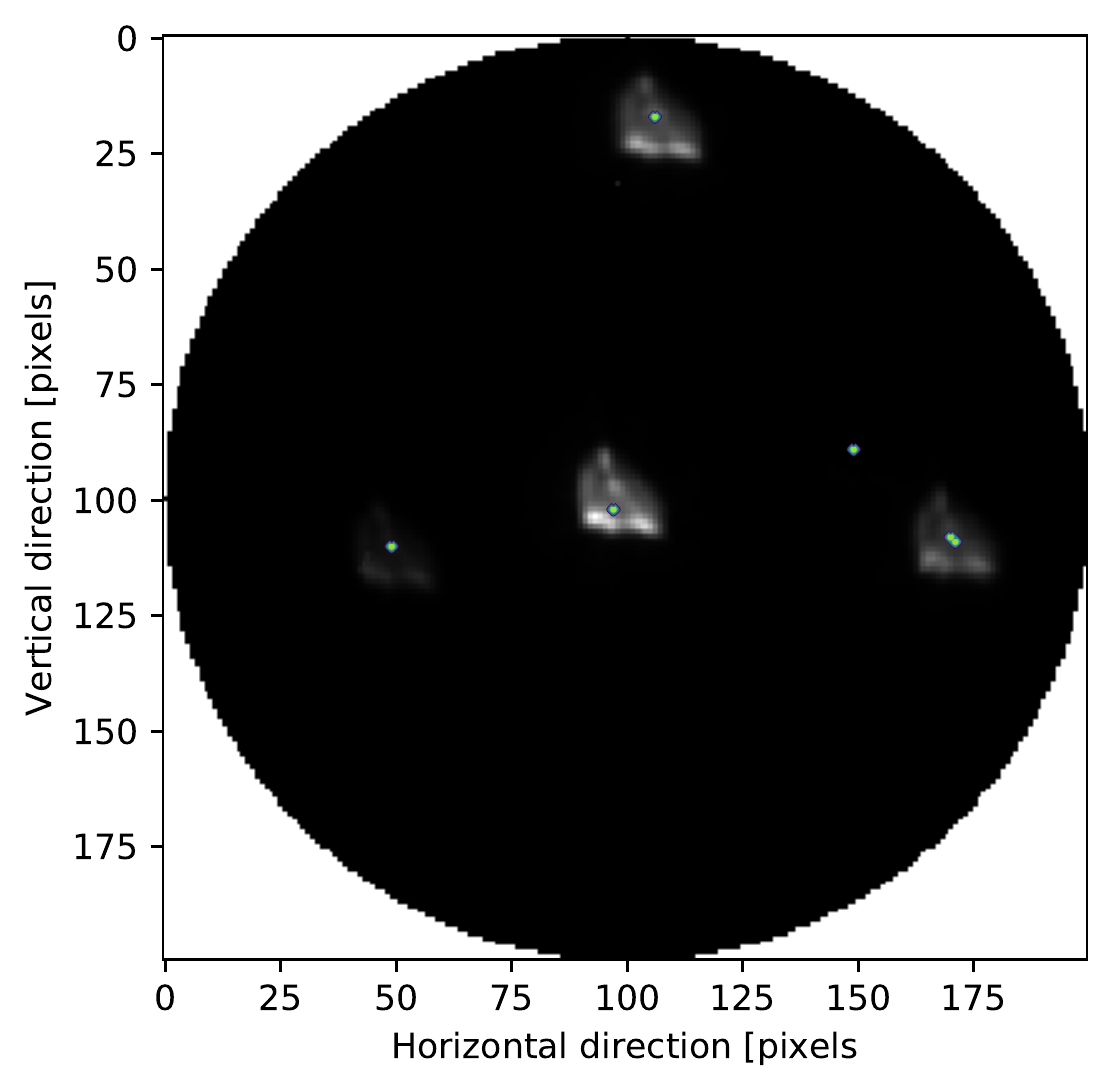}
                \caption{First image of the data set, in the image area, where each point represents the expected position of one star using information from the input star catalogue.}
                \label{fig:all_points_200}
            \end{figure}

        The star catalogue contains many stars, most of which are very faint stars and end up not being actually detected in most CHEOPS observations, as we can see in Fig.~\ref{fig:all_points_200}. Those stars cannot be reliably studied with this pipeline and thus they must be removed from the pool of possible star positions. This can be accomplished with a simple magnitude filter, i.e removing the points from stars with a magnitude higher than 13, as illustrated in Fig.~\ref{fig:all_points}, which is a very rudimentary way of accomplishing it. Since in \textit{CHEOPS} the exposure time depends on the target star there is not a guarantee that this threshold is always valid, since with higher exposure times we could be able to detect stars with an higher magnitude. Thus, we were motivated to use a different approach that was not as sensible to those problems, with image processing techniques.

    \subsubsection{Usage of image processing techniques: contour detection} \label{Sec:dynam_init_track}
        Due to the unique shape of CHEOPS's Point Spread Function, or PSF \citep{off_cheops_pipe}, we attempted to track the stars by detecting them in the images.
        There are many approaches that one can use to track a moving object in images, but feature-based approaches tend to be more robust ones. This approach will be built around image moments, which we can think about as an weighted average of the pixel's intensities that can be used to determine the image's area, centroid and even some information on the image's orientation. A further discussion is beyond the scope of this paper, but an interested reader can refer to \citet{Art:Rochaa}.

        \paragraph*{Shape estimation}
        Using both the zeroth, $m_{00}$, and first degree, $m_{01}$, moments we can estimate the centroid of any given shape, by applying equation~(\ref{Eq:dynam_moments}), in which $X_c$ and $Y_c$ are the coordinates of the image's centroid.
  
        \begin{equation} \label{Eq:dynam_moments}
        \begin{aligned}
        &X_c = \frac{m_{10}}{m_{00}} \\
        &Y_c = \frac{m_{01}}{m_{00}} \\
        \end{aligned}
        \end{equation}

        Such algorithm can be easily implemented using Python's \textit{OpenCV}\footnote{\url{https://opencv.org/}. Accessed: 28/8/2019.} wrapper library, as seen in \citet{cv_moments}. However, before we can apply it, we have to perform some pre-processing steps to the images, so that they match the desired functions inputs. In order to avoid tampering with the photometric results, all changes are made over a copy of the original image that is only used to find the shape of the stars.

        In order to properly use \textit{OpenCV} we must convert our images to a suitable data type. We can convert them to various formats but, the easiest one to convert to, is from one in which each pixel has 16 bits, to one in which each pixel is an 8 bit unsigned integer. Since an 8 bit unsigned integer can only store numbers up to 255, we normalize the image in relation to its brightest point and then scale it up to 255. 

        Now that we have our image in the desired data type, we proceed to apply a binary threshold to the image, as given by equation~(\ref{Eq:bin_thresh}), so that the stars are represented by a value of MaxValue and the background a value of zero, thus facilitating the next step in the process: finding the contours. 

        \begin{equation} \label{Eq:bin_thresh}
        \begin{aligned}
          I_{final}(x,y)=\begin{cases}
            Max Value, & \text{if $I_{original}(x,y) > threshold$}.\\
            0, & \text{otherwise}.
          \end{cases}
        \end{aligned}
        \end{equation}

        Both the contour detection and the moments calculation are, once again, handled by the  \textit{OpenCV} library. As a last step we have to take into account small positives, i.e. a small region of points that barely passed the threshold value and, consequently, its shape is detected. Within this library framework, the contours are returned as a set of coordinates, specifying the 2D coordinates of each point in the contour. If we calculate the area of the PSF, assumed to have a radius of 12 pixels, we can roughly estimate that it should have \textasciitilde 452 pixels in it. Thus, if we discard the masks that have less than 50 points inside them, we can avoid the false positives created by noise or other artifacts on the image, as we do not expect to be able to detect and track (in a reliable way) such small objects.

        If we have one star much brighter than the others due to one of the background stars having an higher magnitude than the target, we might run into problems with the aforementioned normalization routine. As each image is normalized to its maximum flux value, the saturation of one of the stars might push other, fainter, stars below the binary threshold value. To circumvent such problems, we perform an iterative normalization process where, after finding the stars, the N brightest ones, with N being set by the user, are removed and the analysis process is redone.

        \paragraph*{The binary threshold}

        As we expect the detectable stars to be much brighter than the background level, we can set the threshold to a value of 10 which should be more than enough to detect them in the background of the normalized images. If we use  a lower threshold we risk starting to include other noise sources in the images, which is not desired. On the other hand, the usage of this threshold is such that in cases where one of the stars is saturated, or much more brighter than the others, the normalization routine will make the faintest stars dip below the threshold value and thus not detected by the contour-detection algorithms, even though it can clearly be seen in the images, e.g. Fig.~\ref{label-7794}. 

\begin{figure}
            \centering
            \includegraphics[width=8cm]{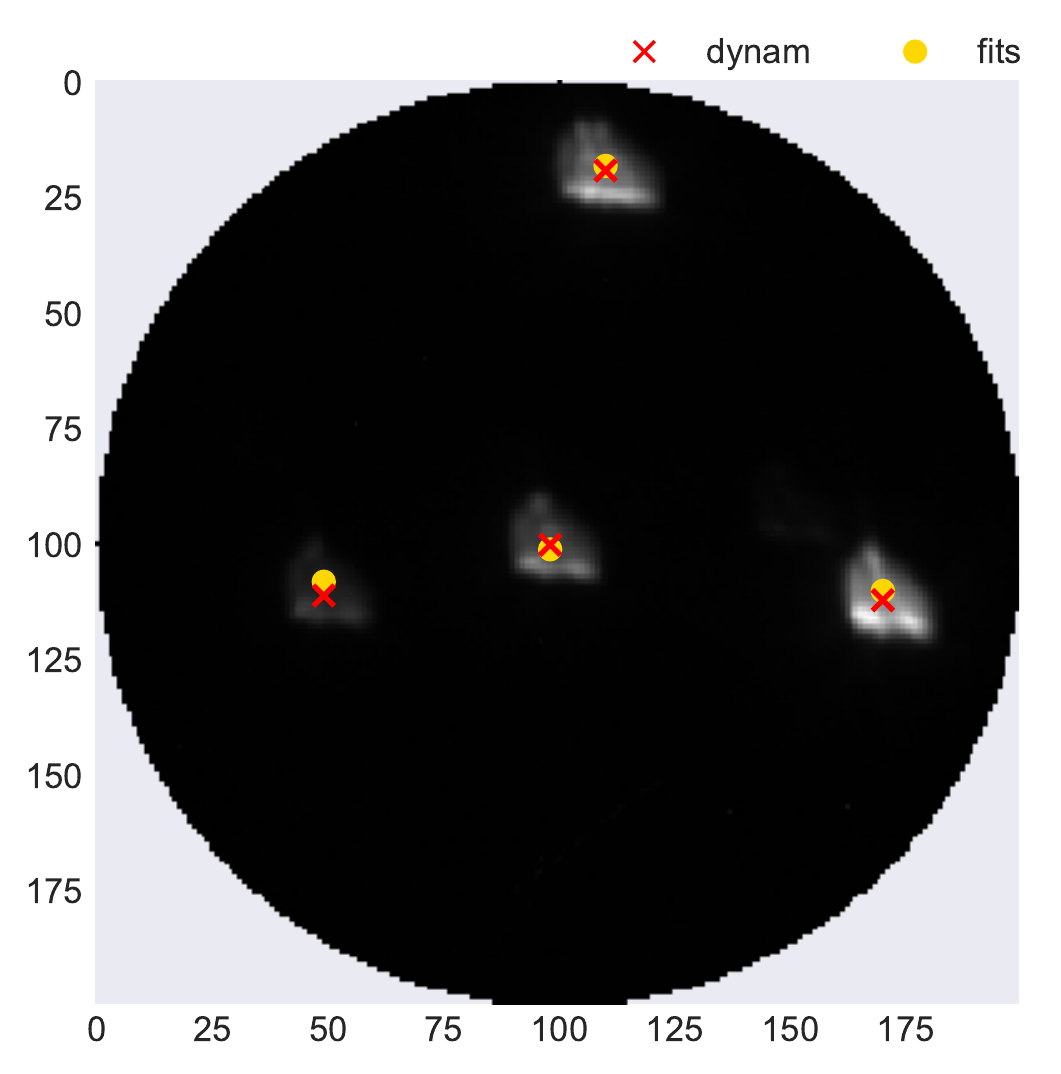}
            \caption{Initial star detection for all the stars using both methods presented so far. With the yellow circles we have the initial detections calculated with the \textit{fits} method and, with the red crosses, the \textit{dynam} method.} 
            \label{fig:all_points}
        \end{figure}

        \paragraph*{Overall characterization of the routine}
        In Fig.~\ref{fig:all_points} we see that when applying this technique, represented with the blue points, we retrieve initial detections very near the ones estimated with the \textit{Star Catalogue}, in yellow.

        Remembering from before, when estimating the position of each star with the \textit{Star Catalogue} we found that there were some faint stars inside the image region. However, when using this method, we cannot find such faint stars, due to either  using a image with less information due to the data type conversion, the fact that the normalization routine is made with the brightest star or maybe the stars are too faint to be reliably detected. 

    \subsection{Tracking the stars}  \label{Sec:star_track}
    
        Due to the satellite's rotation, the CCD is not always in the same orientation but, instead, it is also rotating around the target's direction. To properly study the background stars, we need to be able to track them consistently between images. To accomplish it, we implemented and tested three different methods:
        \begin{itemize}
            \item ``static'': In order to get the points from the i-th image, rotate the ones from the (i-1)-th image, by the rotation angle of the satellite;
            \item ``offsets'': Subtract, from the background stars coordinates obtained with the ``static'' method, the jitter suffered by the central star, when compared against the first image. For the central star, we use the points estimated by the DRP.
            \item ``dynam'': Tracks the stars using image processing, by applying the technique described in Section \ref{Sec:dynam_init_track} to all images.
        \end{itemize}

        Even though all of the approaches are good enough to estimate centroids within the stars, the first two have too much jitter to be able to yield good light curves. Thus, for the background stars, the ``dynam'' method is the best option. Regarding the central star, both the ``offsets'' and the ``dynam'' method present somewhat similar results, depending on \textit{archi}'s configuration, as we shall see later on, in Section~\ref{sec:results}.

    \subsubsection{Shifting the masks} \label{Sec_mask_shift}

        Now that we know the position of the star in each frame, having the masks, Section \ref{Sec:masks}, accompany
        the movements is trivial: we need to calculate the changes in both axis in relation to the
        initial position and, afterwards, shift the initial mask by that amount. This shift is truncated to the pixel size precision, however we can do a re-sampling of the image to increase it, as we shall discuss in Section \ref{Sec:bg_grid}. 

        With the masks moving, we may encounter cases in which the shift is such that part of the
        mask goes outside of the image boundaries, thus picking up “empty” space.  Although the areas outside the image are typically small,
        they can still impact the overall quality of the light curve and introduce errors, mainly due
        to the fact that, in practice, a number of pixels “disappears” for that point in time.
        The mask breaching the image boundaries is expected to occur for stars close to
        the image's edge, where either mask’s shape or small uncertainties in the tracking
        method can lead to this situation.

    \subsection{Photometric masks}\label{Sec:masks}
        Within this section we shall explore the different masks used for photometry. So far we have implemented two masks, a circular binary mask and a binary mask built from the edge of each detected star. The masks are created once and, afterwards, shifted for each image, in order to preserve the number of pixels in it. If we used a non-constant mask, then we would be introducing photometric noise in the light curve, due to having
        a constantly changing number of pixels.

        \subsubsection{Circular Mask}

            Whilst the DRP's mask is a circular binary mask with weighted edges, \textit{archi}'s circular mask is strictly binary. Instead of having this non-binary edge we use a background grid, described in Section \ref{Sec:bg_grid}, to artificially re-sample the image and improve performance. Similarly to the DRP, we have chosen a circular mask to respect the symmetry of the rotating image. The usage of this mask also brings the advantage  of it having a shape that is easily changed, by increasing or decreasing its radius.

        \subsubsection{Shape Mask}

            Once again making use of image processing techniques, we can extract the visible part of the star's PSF, i.e. the line that delimits the star against the background, and use it to create a mask. 

            However, we now need to find a way to change the mask’s size, so it can be optimized. Unlike the circle mask, in which it’s straightforward to increase its size, we now need to find a way to increase it and to quantify such increase. To accomplish this, one can simply add layers of pixels around the shape, until the
            desired size is met. For example, an increase of 1, would add one layer of pixels around the entire mask, and so on.

            \begin{figure}
                \centering
                \includegraphics[width=8cm]{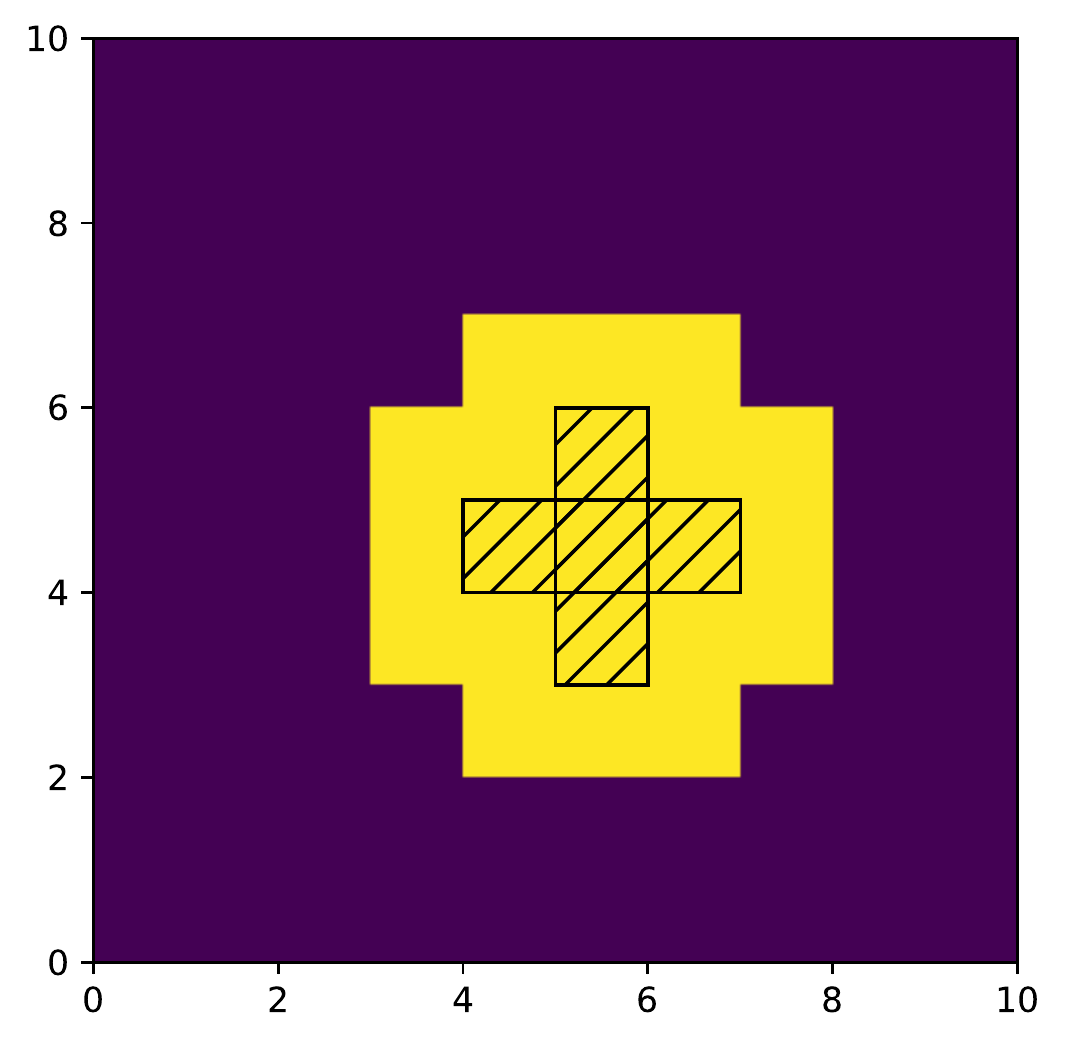}
                \caption{Shape increase method applied to a ``plus'' sign (stripped boxes), that was increased by one pixel.} 
                \label{fig:cross_increase}
            \end{figure}

            In Fig.~\ref{fig:cross_increase} we can see the method working, on a mask resembling a ``plus'' sign, adding a layer of 1 pixel around its edges. It's also noticeable the flaw in the increase of the mask: Simply adding a layer of 1 pixel around a mask, does not maintain its proportions. However, despite not being able to properly increase the mask whilst maintaining the original shape, this will allow us to have a finer control of the mask size during the optimization routine, where we attempt to find the optimal mask size.

        \subsubsection{Finding the optimal mask}

        After determining the initial masks, they are either equal to the star’s shape or a circle
        with a previously defined radius. However, we want to find the best mask for each star,
        i.e. the one that has the largest area of the star's PSF, without picking up too much noise from the background. To do so, we change the mask's size until we find the minimal value of the noise metric, Section \ref{Sec:CDPP}. During this process, we expect the noise to decrease until it eventually hits a minimum value and, afterwards, starts increasing, as the mask starts covering a greater portion of the sky.
        
        If during the optimization routine a mask finds itself occupying regions outside the image, we have a trade-off between the number of points for the light curve and the global noise in it. We gave preference to a higher number of points in the light curve to try to better assess the global behaviour instead of beneficing the precision in each point of smaller sampled light curve.
    \subsection{Background grid} \label{Sec:bg_grid}
        
        Due to the usage of shape-based algorithms for the masks and star detection, alongside the integer conversions that must be made to convert positions into grid coordinates, the movements on the image are limited to the grid’s nodes. Thus, ideally, we would like to have a image with more resolution than the one that comes out of \textit{CHEOPS} CCD.

        When working with images, we do not have a continuous
        surface and, instead, we should think of it as a equally spaced grid, in which each node
        represents a pixel. Thus, if we wish to link the determined coordinates of a star with a given pixel in the image,
        we must make an approximation. Once again thinking on the pixel grid, all detections the lie within the grid node 100 and 101, are considered to be in the same pixel. Consequently, the mask is placed in the pixel that corresponds to the integer part of the estimated location.

        \paragraph*{Re-sampling the images}
        By increasing the number of nodes in
        our grid between any two pixels, we could approximate the coordinates to a node much
        closer to their actual value.  To do so we can artificially re-sample the images, with a positive scaling factor, defined in equation (~\ref{Cha_photom_scaling_f}), greater than one.

        \begin{equation} \label{Cha_photom_scaling_f}
             scaling\ factor = \frac{ N_{increased} }{N_{original}}
        \end{equation}

        , where $N_{increased}$ is the number of points in each side from the increased image, which can be chosen by the pipeline user, and $N_{original}$ is the number of points in each side of the original image.

        Furthermore, we wish that a single pixel in the original image is transformed into a block of pixels, in the re-sampled image. In order to facilitate the conversion between images, we shall only use odd values for the scaling factor.

        To finalize, there is a detail that we must keep in mind: we can see the increase in the grid size, as an increase in the correlated points, i.e. if one pixel is transformed into 9, then those 9 pixels , before normalization, are equal to the first one, as well as all the errors associated with that pixel. To avoid introducing flux into the images, we normalize the block of pixels, so that its sum is equal to the original level. We confirmed conservation of the flux, up to machine precision, on the re-sampled grid.

    \subsection{Uncertainties} \label{Sec:archi_uncerts}

        In the DRP the calculation of uncertainties in the light curve is made through Equation \ref{Cha1:uncertainties}:

        \begin{equation} \label{Cha1:uncertainties}
        \begin{aligned}
            &Err = \sqrt{Flux + bg + N_{pix}N_{stack}(gain * ron)^2 + dark*t_{exp}N_{pix} } \\
            & bg = background*N_{pix}*t_{exp}
        \end{aligned}
        \end{equation}

        where Err is the uncertainty for a given point, Flux is the flux of the corresponding point, $N_{pix}$ the number of pixels inside the mask, \textit{$N_{stack}$} is the number of stacked images, \textit{gain} is the gain from the digital conversion process that occurs in the CCD, \textit{ron} is the read out noise, \textit{dark} is the dark current, $t_{exp}$ is the exposure time for each image and, lastly, \textit{background} is the flux from background objects.. It's important to note that this equation does not take into account the jitter of the spacecraft, since adding it is not trivial. In \textit{archi} we shall also use this equation to estimate the uncertainties, although some care is needed.

        First of all, the parameters stored in the DRP outputs are calculated over a 200x200 region, giving us a per-pixel value of each component, as we shall see in this Section. However, when using a background grid, the number of pixels inside the masks will be inflated, when compared to the cases in which the background grid is not in use. Thus, if we do not correct this inflation, we will have an overestimation of the uncertainties.

        In order to correct this effect we have to  convert the number of pixels from the increased grid to the normal one, which is accomplished with Eq.~\ref{eq:change_grid_nmb_pix}.

        \begin{equation}\label{eq:change_grid_nmb_pix}
            corrected\ size = \frac{total\ size}{(scaling\ factor)^2}
        \end{equation}
        where \textit{corrected size} is the corrected number of pixels inside the mask and \textit{total size} the total number of pixels.

        Another consideration that must be taken into account is that both the background and dark values are stored for the entire region of the DRP's aperture, instead of being a per pixel value.  Since the DRP's mask is not a completely binary mask, as near the edges the pixels are weighted, it will not be possible to retrieve the exact values. As an approximation we shall consider the DRP's mask to be a circular binary mask, thus allowing us to compute the per-pixel values, as we shall now see.

        \subsubsection{Background} \label{Cha3:uncert_bkg}

            The background stored in the DRP outputs is not the calculated value but, instead, it is re-scaled by mask's size, as given by Eq.(~\ref{Eq:actual_bg}).

            \begin{equation} \label{Eq:actual_bg}
                stored\_bg =  background * t_{exp}*N_{pix}
            \end{equation}

            With this knowledge, we want to extract the background per pixel, to calculate the uncertainties for our mask of choice. One aspect to take into account, would be the fact that the DRP's background calculation is made over the image outside a region delimited around the target star, i.e. in the background value we can find contributions from the background stars. We could try to improve the background calculation with the masks and star tracking techniques so far described but, we would be introducing errors due to the movement throughout the CCD pixels and, some parts of the DRP would need to be re-implemented to allow us to work with images without background correction already applied on. Furthermore, as in the DRP \cite{off_cheops_pipe} the background correction routine already takes into account contaminating stars, we can use it to calculate the errors on our light curves. Thus, if we divide the stored\_bg by the number of pixels inside the DRP's mask, we get an estimation of the number of dark photons during the exposure time (actually exposure+reading time).

        \subsubsection{Dark}
            As CHEOPS has a small field of view and, consequently, the background stars are relatively close to the target, we can make a fair assumption that the dark value, estimated by the DRP for the central target, should be reliable also for these close by background stars. Furthermore, the typical dark current has a low value \cite{off_cheops_pipe}.
        
            Similarly to the background, the \textit{dark} stored in the outputs is only calculated for the region near the central star, using the image outside the applied mask. 
           
            The actual values are stored in temporary files that are not available in the DRP's processed data. Thus, to have this information, we would have to recalculate it.  Even though the \textit{dark} value is not stored, in DRP's outputs we can find the dark component in the uncertainty calculation, given by equation  (~\ref{Eq:actual_dark}).

            Since we know the radius of the circular mask used for our data set, we can simply divide this stored value by the number of pixels inside the mask, and thus have a rough estimation of the \textit{dark}. However, DRP makes use of 4 different masks and, if we are not careful, nearby stars can impact its value. Thus, in an attempt to minimize the contaminations, we can calculate the median of the estimated \textit{dark} for each one of the 4 DRP's apertures, and regard it as the \textit{dark} value.

            \begin{equation} \label{Eq:actual_dark}
                stored\_dark = dark *  t_{exp}*N_{pix}
            \end{equation}

            The downside of this method is that it assumes that we have an uniform dark, that is equal for both the target and the background stars, which may not hold as true. However, since the contribution from photon noise is a few orders of magnitude higher than the one from the dark, those imperfections should not be very impactful. More so, the field of view is very small and therefore we expect that the dark will not change much in the image, specially in the \textit{subarray} image, which is just a small piece of the CCD.

\subsection{Noise metric - Combined Differential Photometric Precision} \label{Sec:CDPP}

    In order to estimate the noise in the light curves, an adaptation was made to the algorithm applied in NASA's Kepler mission: Combined Differential Photometry Precision, CDPP, \citep{Art:Christiansen2012}. 

    In \citet{Art:Christiansen2012} words: ``\textit{A CDPP of 20 ppm for 3-hr transit duration indicates that a 3-hr
    transit of depth 20 parts per million (ppm) would be expected to
    have a signal-to-noise ratio (S/N) of 1, and hence produce a signal of strength 1 $\sigma$ on average}'', which is the ideal metric to quantify the noise existent on the light curve and, it can be interpreted as the effective noise seen by a transit pulse \citep{Art:Christiansen2012}. In its calculation, the important factor is the near-term trend changes in brightness instead of the long-term ones \citep{Art:Koch2010}.

    For further details on the algorithm itself, one can refer to \citet{Art:Jenkins2010,Art:Christiansen2012}, where the algorithm is properly introduced and discussed. However, for a brief introduction on the methodology behind it, I will now refer to one of its many adaptations, in \citet{Art:Luger2016}:

        \begin{itemize}
        \item Start by passing a 2 day quadratic \textit{Savitsky-Golay} (SavGol) filter to the flux, which should be more than enough to avoid fitting the transits in K2 Light Curves;
        \item Remove 5 $\sigma$ outliers;
        \item Divide the data into chunks, which have data from a time interval equal to the integration time. Afterwards the standard deviation is calculated  for each one of them;
        \item Take the median of the standard deviations and divide by the square root of the number of points inside each chunk, obtaining the desired photometric precision.
    \end{itemize}

    The official DRP reports the usage of a modified CDPP algorithm, better described in \citet{off_cheops_pipe}, which does not use any kind of detrending nor filtering within the metrics, also capable of taking into account eventual gaps in the data. The same algorithm was used in archi, to calculate the CDPP. To estimate the errors in this measurement, we built a 68 \% confidence interval of the standard deviation, across the chunks, with the 16th and 84th percentile. Thus, with this metric, we can evaluate the quality of the raw light curves.

\section{Results} \label{sec:results}

    Using the presented methods, we built \textit{archi: An expansion foR the CHeops mission pipelIne }, an open source python package \footnote{\url{https://github.com/Kamuish/archi}}, that works on top of the DRP outputs, allowing one to extract information from all of the stars present in \textit{CHEOPS} observations. In the \textit{github} repository one can find more information on how to use the library, alongside practical examples and, in \citet{MSc:THESIS:andre_m_silva}, the pipeline architecture is thoroughly discussed.

    Within this Section, we shall compare the different approaches and options in archi, i.e. the mask to be used, the size of the background grid, the initial detection method and the star tracking methods. This comparison will be made through the photometric precision, for a timescale of 30 minutes. As we still do not have real data from the mission, we made use of \textit{CHEOPS} official simulator tool \citep{Art:cheops_Sim} to create simulated data sets.

    \paragraph*{Simulated Datasets}
        In order to study the behavior of \textit{archi} we will use three different batches of simulated datasets. Data Set A has three stars rotating around the target star, all with different magnitudes. Data Set B1, B2 and B3 all have four stars, with the background ones having equal magnitudes in all three data sets. The only difference between them is the magnitude of the target star that is different to have different exposure times. Lastly, Data Set C has two visible stars rotating around the target star, all with the same magnitude. In, in Appendix ~\ref{APP:datasets} one can find more information on the different simulated datasets.

    \paragraph*{Naming conventions}
        To simplify the linking of light curves to actual stars, each star is named using the following convention: ``Star <index> (separation, $\Delta mag$)'' where index is an integer linked to the closeness to the centre of the image, separation is the distance from the target to the star and  $\Delta mag$ is the difference of magnitudes between the star and target. Following this convention, the central star has index zero, the closest star to the target has index one, with the pattern being kept for all other stars. In Fig. ~(\ref{init_pos_stars}) we can see the name of each star, alongside their initial position in the image.

        Lastly, from this point onwards, we shall set the convention of  <mask type> - <Initial detection> - <star tracking> to refer to the combination of methods that was used to obtain a given light curve. 

    \subsection{Light curve extraction}\label{Sec:compare_archi_methods}   

        Now that we have seen the different methods implemented in \textit{archi}, we shall now characterize it. To do so, we shall start by looking at Data set A, comparing the different combinations of star-tracking methods, presented in Section~\ref{Sec:methods}. 

        In the first place, we will see how the different combinations of masks and star tracking interact with each other. We have tested all combinations of initial detection methods and star tracking methods and decided to only show the ones that can achieve the best results and compare those against each other. Thus we shall only use the \textit{dynam} initial detection method and the \textit{dynam} and \textit{offsets} star tracking methods.

        \subsubsection*{Target star analysis}
\begin{figure*}
            \centering
            \includegraphics[width=15cm]{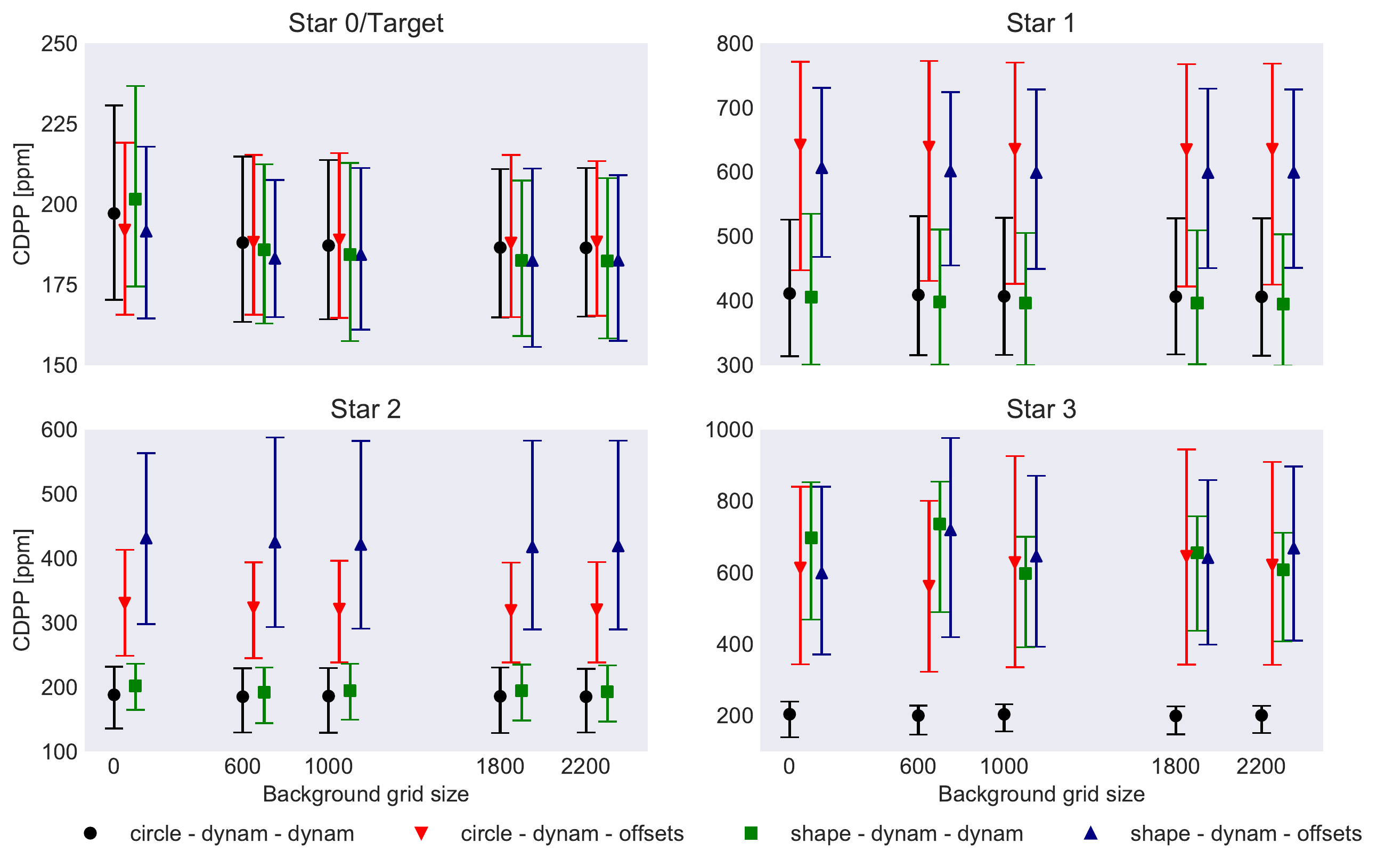}
            \caption{\label{comp_all_A}Comparison between the CDPP of the light curves produced with the \textit{shape} and \textit{circle} mask, for Data Set A.}
        \end{figure*}

        We can start by looking at the target star in Fig.~\ref{comp_all_A} that, in a generalized way, has both masks yielding noise values close to each other, assuming that the same star tracking method is used. Even between both star tracking methods, we fail to find big differences between the worst and best noise value. It's also noteworthy that the increase of the background grid allows us to diminish the noise in the light curves. However, as one should expect, this improvement is limited, and grids bigger than 600, corresponding to a scaling factor of at least 3 times the original size of the image, do not reduce significantly the noise, despite the associated increase in the computational cost.  Instead, from this point onwards, we find some small fluctuations in the noise level, which is a sign that we are close to the optimal conditions for the pipeline to work.

        For this star, the error bars in the CDPP values stay relatively constant, without any visible changes neither between different methods nor with changes in the background grid in use.

    \subsubsection*{Benchmarks against the DRP}
        Now that we have seen the performance of \textit{archi} for the target star we can compare it against the DRP's OPTIMAL light curve.

        \begin{figure}
        \centering
                \includegraphics[width=9cm]{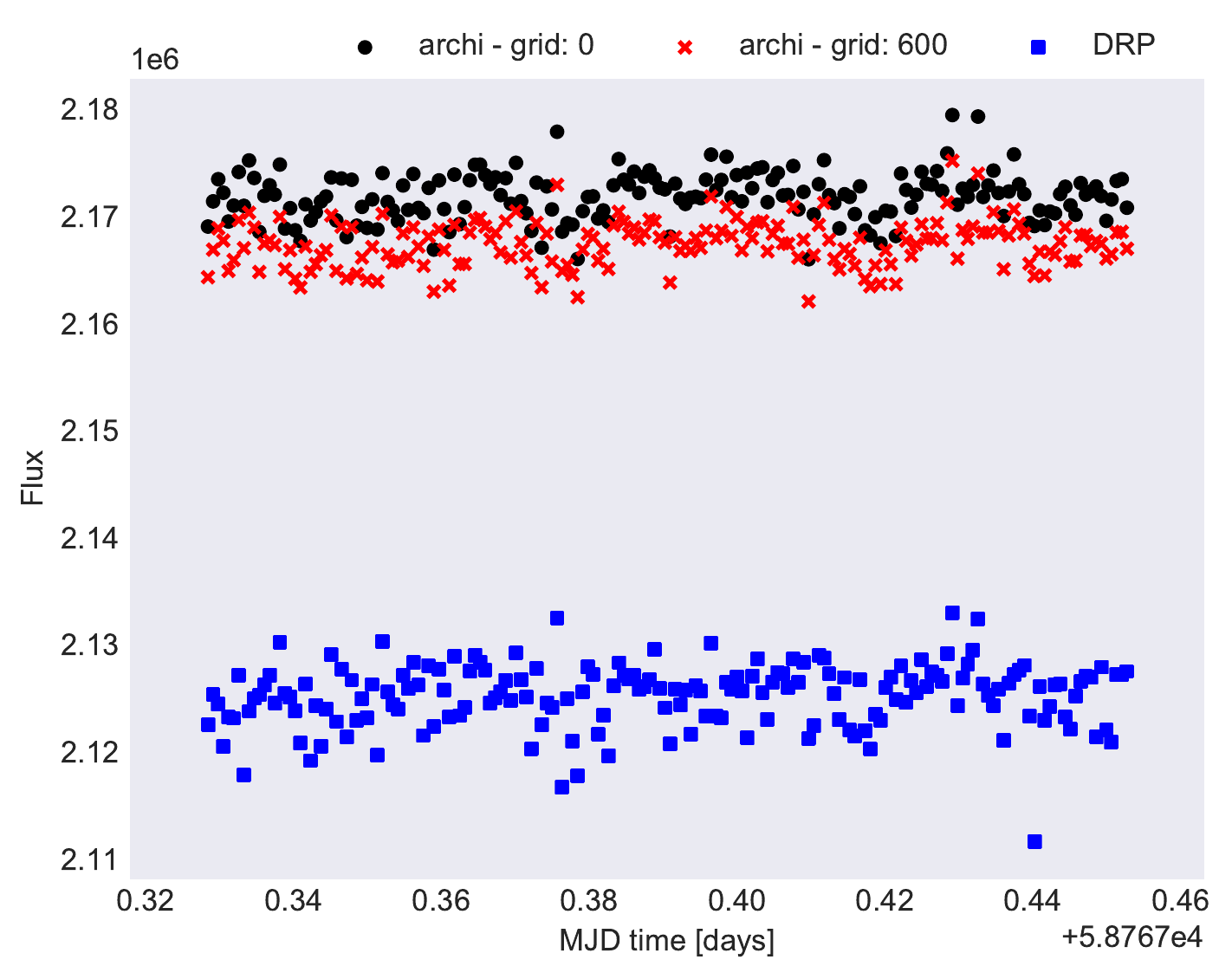}
                \caption{\label{DRP_compare}Light curves extracted, from the target star, without a background grid, in the black dots, with a grid of 600 in red crosses and, in blue squares, DRP's \textit{OPTIMAL} light curve. Both of \textit{archi}'s light curves were obtained with a shape - dynam - offsets combination. In Table ~\ref{tab:archi_drp_comp} we can find the noises associated with each light curve. Data Set A was used for this comparison and, in Table~\ref{tab:archi_drp_comp} we can find the photometric precision alongside the corresponding error bar, for each light curve.}

        \end{figure}
        \renewcommand{\arraystretch}{1.5}
        \begin{table}
            \centering
            \caption{Comparison of the CDPP for the light curve extracted, with \textit{archi} and the DRP, from the target star, while using Data Set A.}
            \label{tab:archi_drp_comp}
            \begin{tabular}{lcc} 
            \hline 
            Pipeline & Noise [ppm] \\
                \hline 

            DRP & $254^{+48.0}_{-43.3}$ \\
            archi - grid 0  & $192^{+26.2}_{-27.1}$ \\
            archi - grid 600 & $183^{+24.3}_{-18.3}$
            \end{tabular}
        \end{table}

        From Fig.~\ref{DRP_compare} and Table ~\ref{tab:archi_drp_comp} we see that the DRP produces a light curve with a lower flux value, albeit with an higher noise metric and a broader confidence interval. The higher error bars suggest that the DRP light curve is more affected by the systematics introduced by the rotation, since the points are more dispersed within each chunk used for the CDPP calculation.

        In all comparisons made during this work, we have found that \textit{archi} was consistently able to produce curves with a lower CDPP. Although, it's important to note that the DRP is still under development and that this fact may not hold for later versions of it.

        \subsubsection*{Background stars analysis}

        Contrasting what was found for the target star, when looking at the background stars, in Fig.~\ref{comp_all_A}, we find that the star tracking methods return results far apart from each other. 

        For the first two stars, Star 1 ($\ang{;;5}$,+1) and Star 2 ($\ang{;;7}$,-1) , we see that both masks give similar results, with the main difference laying in the star tracking method. The \textit{dynam} method reveals itself to be the superior alternative to track the rotating stars. If we now pay attention to the noise evolution with the background grid, we see that it stays almost constant for them all, without improving or worsening our results. The non-stationary nature of those stars translates into errors in the mask position and mask shape itself, which have not been counteracted by the increased resolution of the image. Within the background stars, we should also pay some attention to Star 3 ($\ang{;;8}$,-0.5) , the closest to the image's border. In this one, the mask size is limited due to its closeness to the edge, thus impacting our ability to obtain light curves with a photometric precision near the ones obtained for the other stars. Furthermore, for Star 2 ($\ang{;;7}$,-1) and Star 3 ($\ang{;;8}$,-0.5)  we can clearly see that the error bars are substantially smaller for the points obtained with the \textit{dynam} star tracking method.

        Following that, we also see that the \textit{circle} mask is much better than the \textit{shape} one, for Star 3 ($\ang{;;8}$,-0.5), assuming \textit{dynam} star tracking methods. We believe that this occurs due to the \textit{shape} mask having an irregular shape and thus it's easier for it to leave the image region, leading to a stricter constriction in its size, when compared with the circular mask.

        \subsubsection*{Analysis of the optimal methods}

        The last characterization made with this data set will be through the methods that were able to produce the best light curves. As we could not find a single optimal combination for both the target and the background stars, we used the optimal for each of the cases, i.e. a combination of shape-dynam-offsets for the target star and a combination of circle-dynam-dynam for the background ones, as \textit{archi} can use different methods, independently, for both the target and background stars. This comes from the fact that the target star remains centered and sampled in the same pixels, while the background stars rotate and travel throughout many different pixels. Furthermore, we also tested with a background grid with 600 points and without it, in an attempt to find artifacts introduced by its usage, whilst maintaining the same set of combinations.

\begin{figure*}
        \centering
                \includegraphics[width=15cm]{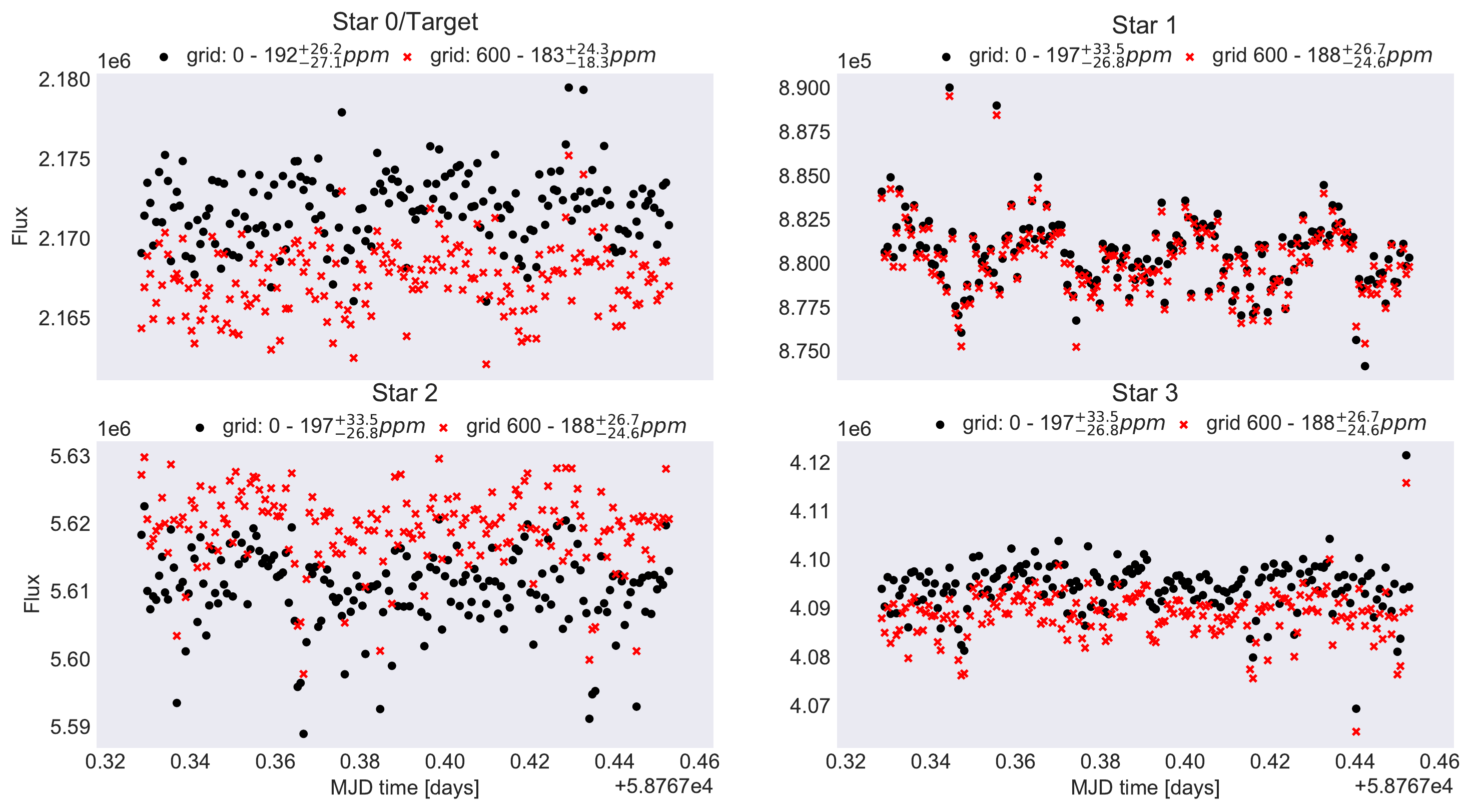}
                \caption{\label{A_best_LCS_image}Light curves extracted, for Data Set A, without a background grid, in black, and with a grid of 600, in red. For the target star we used a combination of shape-dynam-offsets and, for the background ones a combination of circle-dynam-dynam.}
        \end{figure*}

        In Fig. \ref{A_best_LCS_image} we find that the background grid in use does not impact, significantly, the flux level, although it manages to reduce the measured noise. Interestingly, we can find a sinusoidal signal in the background light curves, most noticeable on Star 1 ($\ang{;;5}$,+1).  In \citet{MSc:THESIS:andre_m_silva}, we found that this signal was induced by the irregular shape of each star's Point Spread Function, which was crossed by the background stars during their rotation. In this image we cannot find the same signal for Star 2 ($\ang{;;7}$,-1), due to it being brighter than the target star, and thus the PSF's impact is not as prominent.

        Furthermore, when this contamination is not as present we find that the CDPP shows values close to the ones exhibited on the central star, with similar confidence intervals.

    \subsection{Impact of the target star's magnitude} \label{Sec:impact_exposure}

        Since the exposure time is set to be the ideal one for the central, target star, we can find cases in which the background stars are brighter than the target and thus, they might impact negatively the star detection algorithms. To see how this can impact our capability of detecting the background stars, we shall now look into Data Set B.

\begin{figure*}
        \centering
            \mbox{
            \subfigure[\label{label-7796}Detected stars on Data Set A.]{\includegraphics[width=5cm]{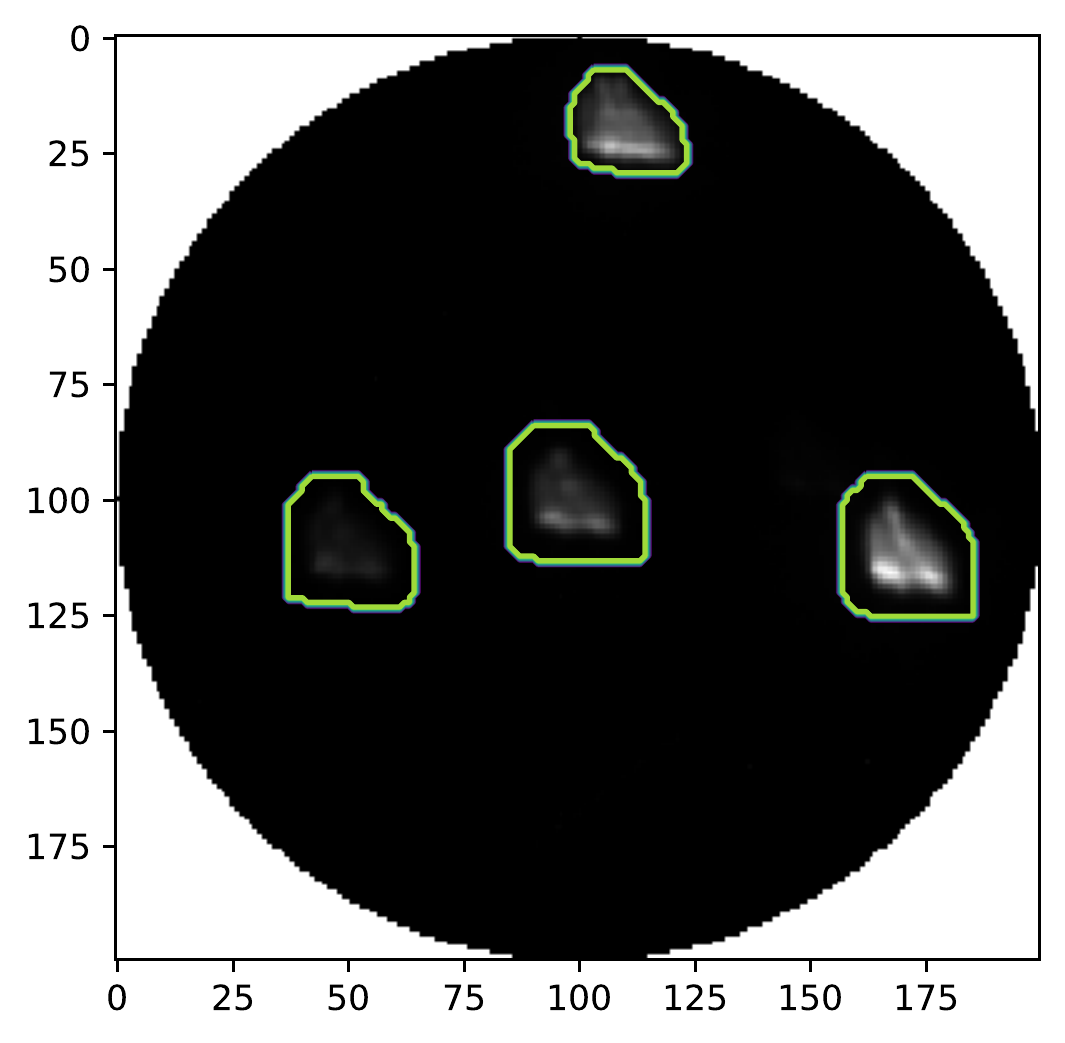}}\quad

            \subfigure[\label{label-7793}Detected stars on Data Set B1.]{\includegraphics[width=5cm]{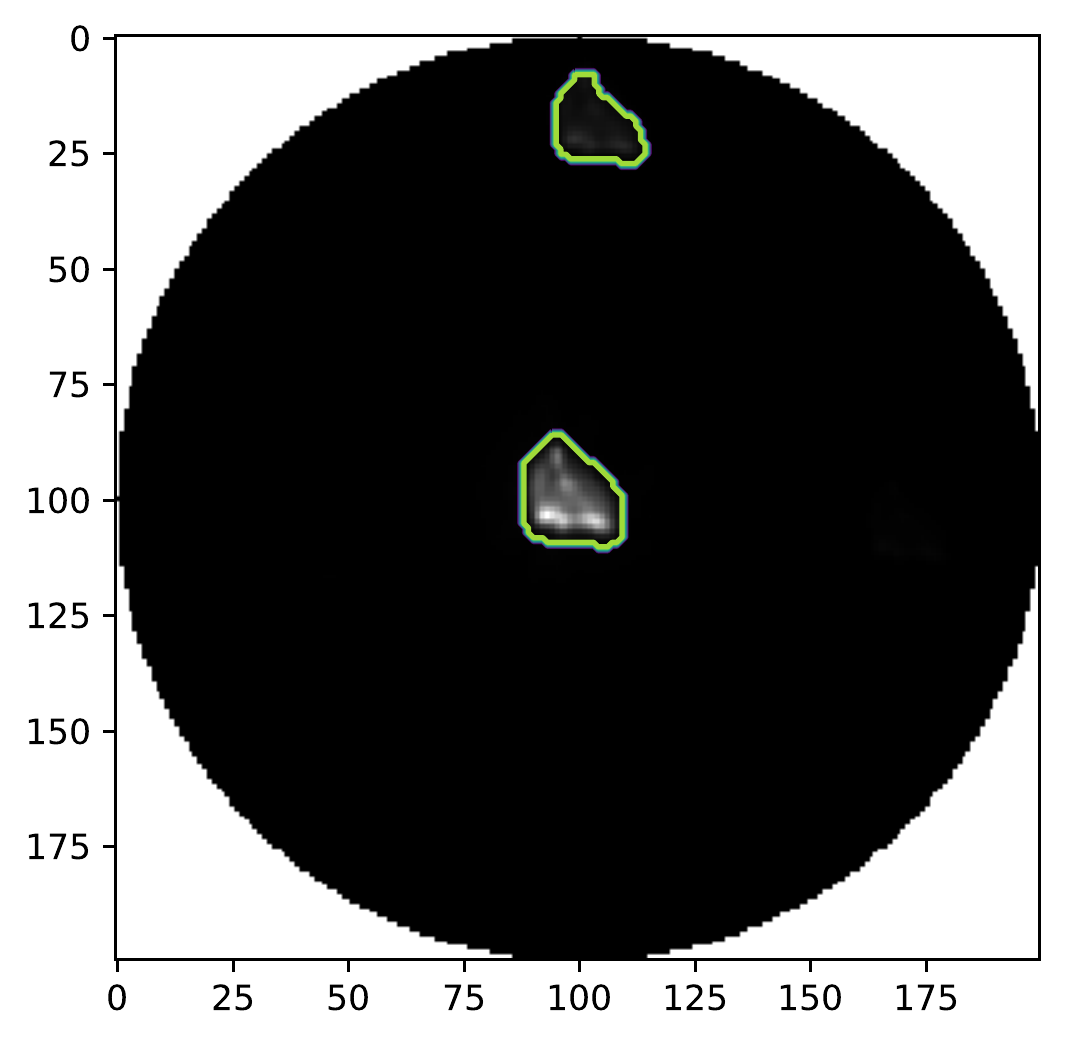}}\quad
            }
            \mbox{
            \subfigure[\label{label-7792}Detected stars on Data Set B2.]{\includegraphics[width=5cm]{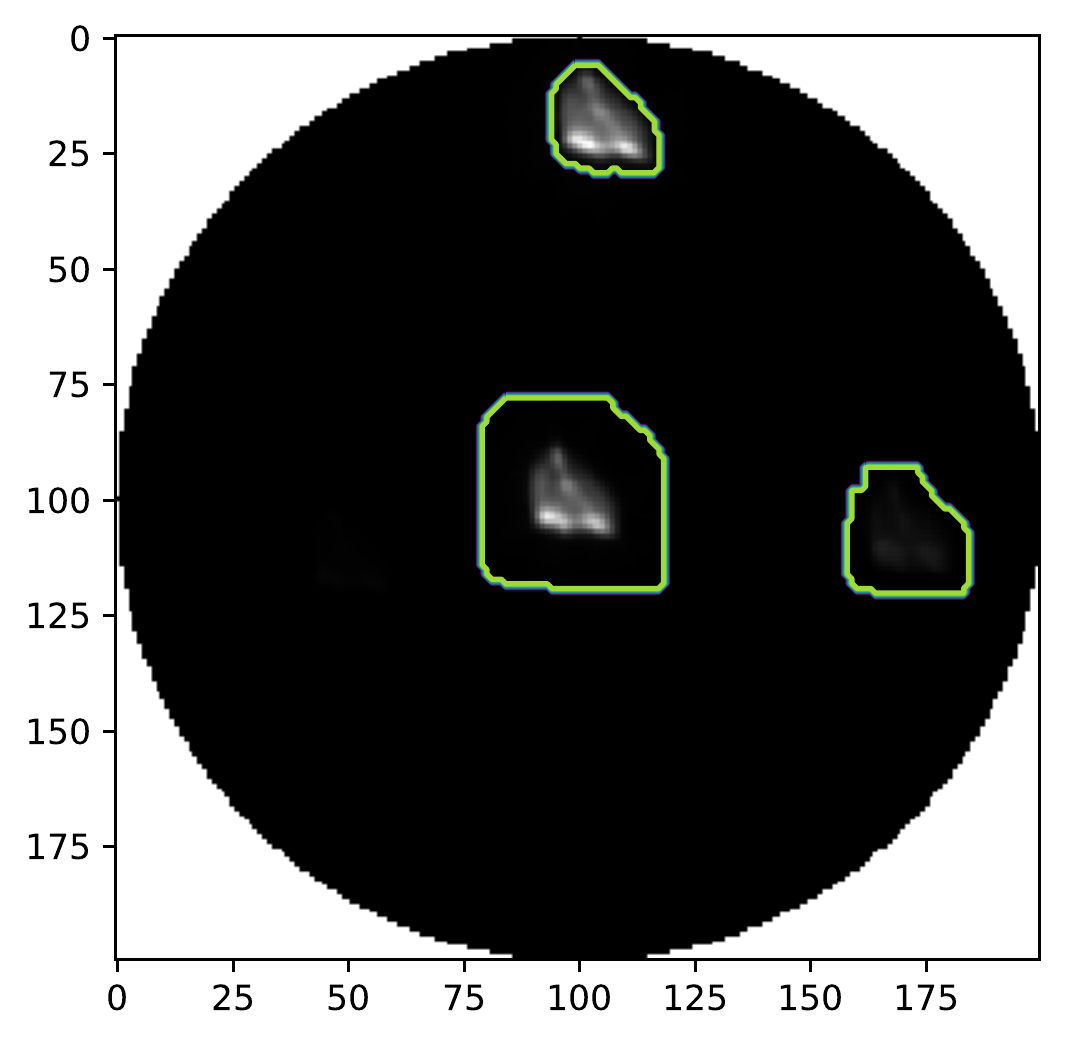}}\quad

             \subfigure[\label{label-7794}Detected stars on Data Set B3.]{\includegraphics[width=5cm]{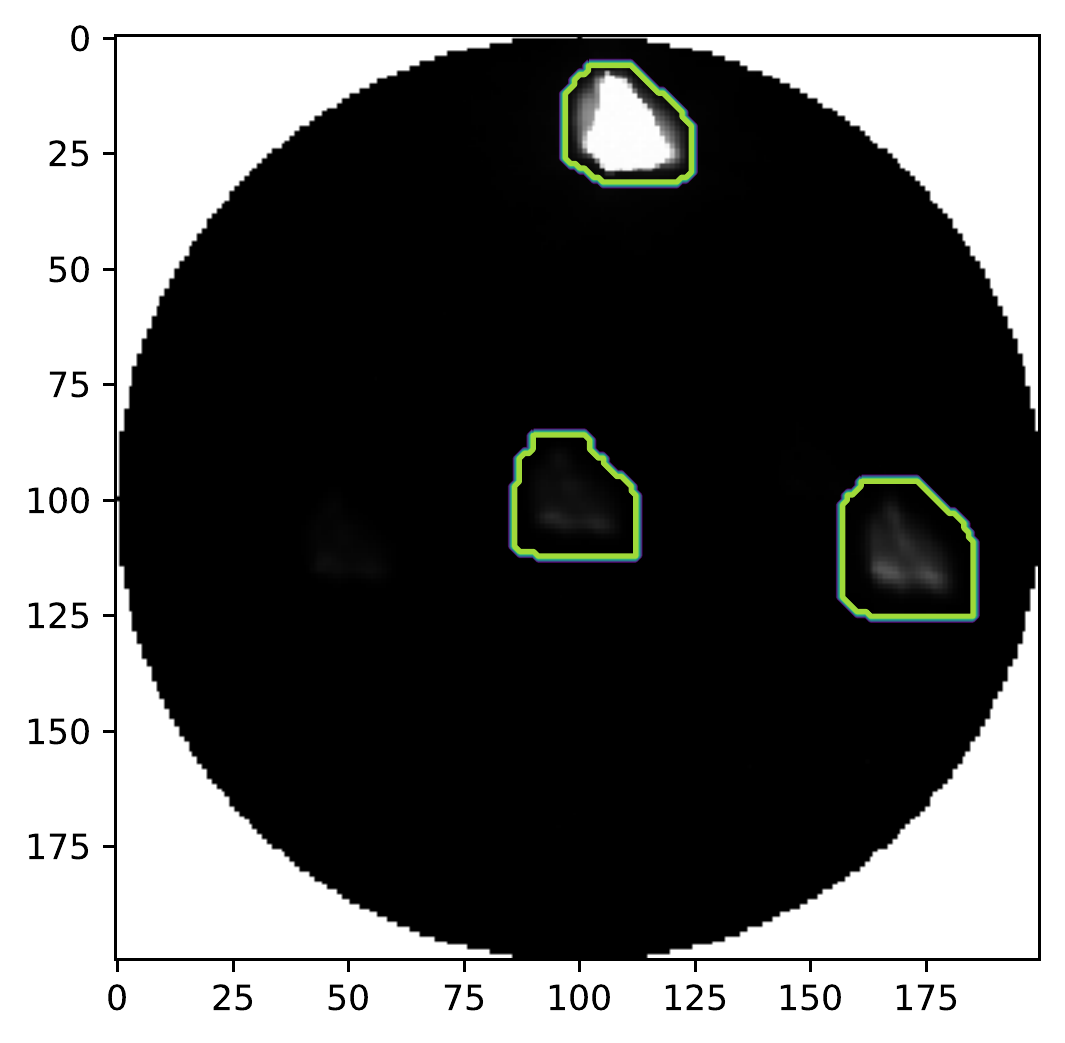}}\quad
  
            }
        \caption{\label{B_before_iteration}Optimal masks for data set A and B. With Data Set B we can see the limitations of the basic normalization routine, as the faintest stars are not found, due to the presence of brighter stars in the image.}
        \end{figure*}

        Starting with Data Set B1, in Fig.~\ref{label-7793}, the case in which the target star is the brightest star, the exposure time is shorter, thus not allowing to collect as much photons for the background stars which leads to us not being able to detect them. The sole exception is the faint star found above the central star, since it's the brightest background star. If we now use a target star not as bright but maintain the magnitudes of the background stars, as seen in Fig.~\ref{label-7792}, we are capable of detecting another star, although it's barely visible.

        Lastly, if we reduce even further the central star's magnitude  to 12 mag, in Fig.~\ref{label-7794}, the closest star to the target is still not detected. The long exposure time needed to observe the target star leads to the saturation of Star 3 ($\ang{;;8}$,-3) on top of the image. As the nearest star to the target, Star 1 ($\ang{;;5}$,-1), is equal to the the one in Fig.~\ref{label-7792}, from data set A, we would also expect to detect it. However, that is not the case due to a limitation of the default image pre-processing normalization routine.

        However, if we use an iterative normalization routine, as described in Section~\ref{Sec:dynam_init_track}, we are able to detect the fainter stars, with the downside of an higher computational cost.  Even though this cost is not extremely evident for smaller background grids, it can quickly ramp up for larger ones, as we are doubling the number of times that we must apply our analysis routine. 

        \begin{figure*}
        \centering
                \includegraphics[width=5cm]{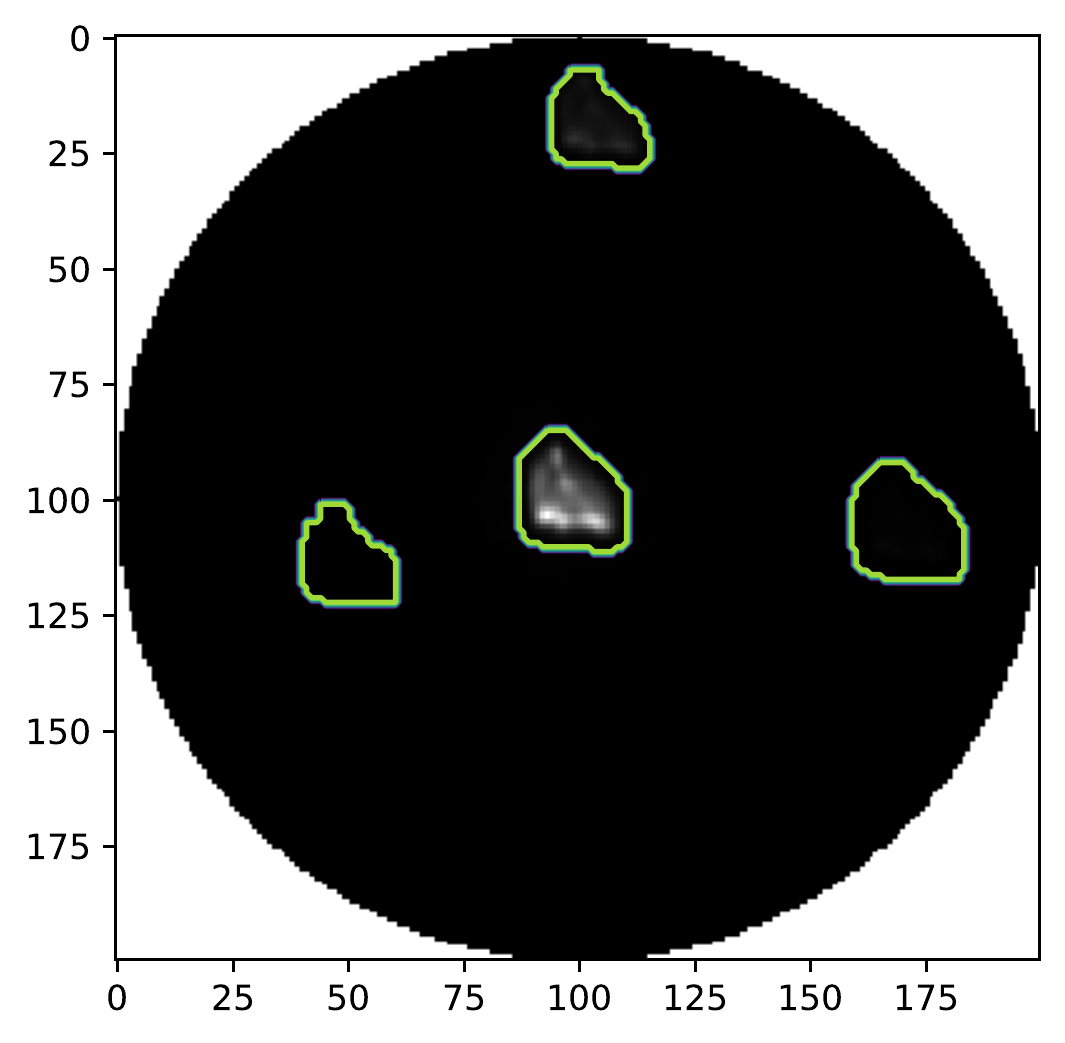}
                \includegraphics[width=5cm]{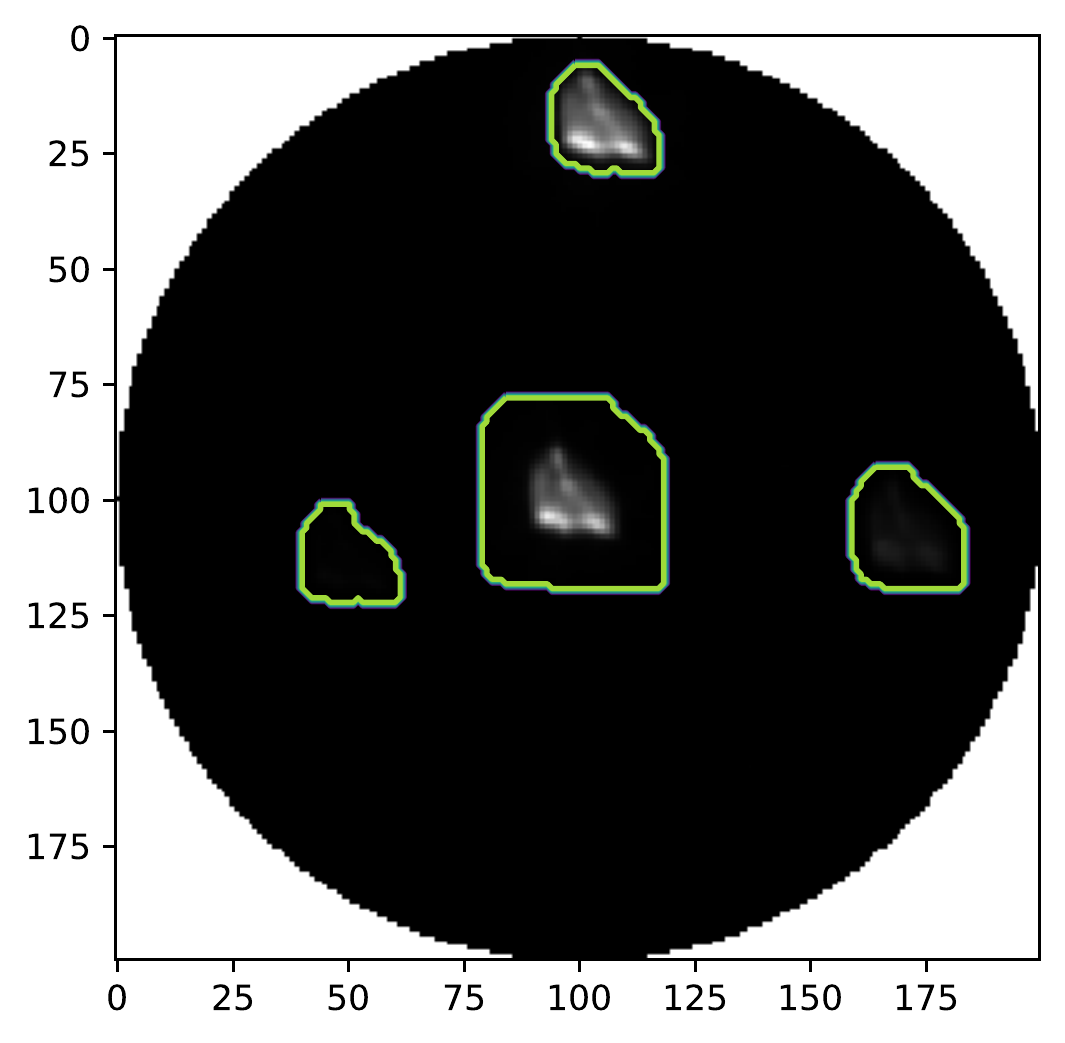}
                \includegraphics[width=5cm]{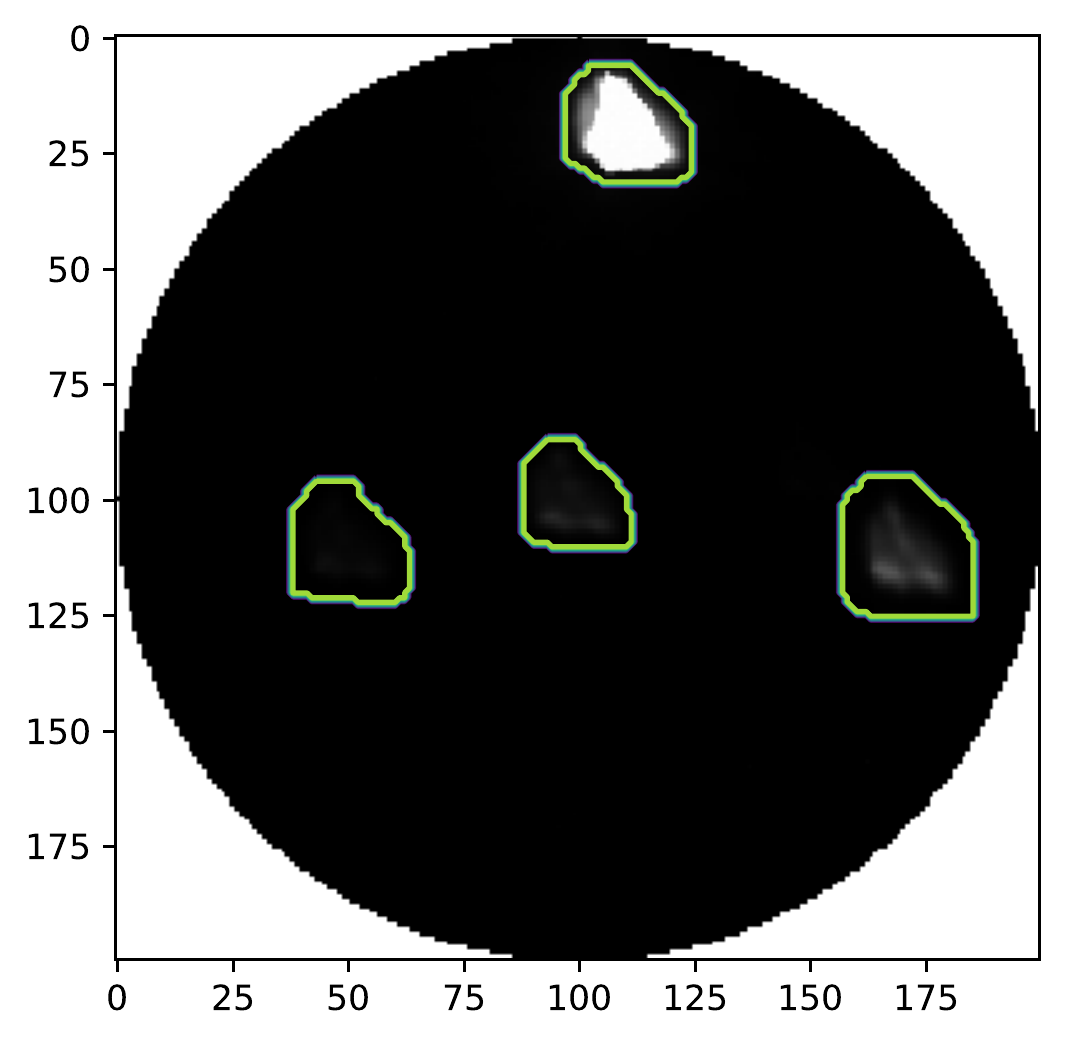}
                \caption{\label{B_after_iteration}Optimal masks for Data Set C, with an iterative normalization routine. \textbf{Left:} Data Set B1, with 2 iterations; \textbf{Middle:} Data Set B2, with 2 iterations; \textbf{Right:} Data Set B3, with 1 iteration.}
        \end{figure*}

        From Figure \ref{B_after_iteration} we see that, with the iterative normalization routine, we are able to detect the previously undetected stars, seen in Figure \ref{B_before_iteration}. However, as we are removing regions of the image, there is always a chance that we will leave some residual pattern, that can be detected as a star. On top of that, the over-usage of this iteration process, may lead to cases where some regions of the background start being above the binary threshold in use. As the faintest stars are able to detected, is unlikely that stars not accounted for will contribute to signal found on the target star.

        Attempts to work with the logarithm of the images, to avoid the double analysis of the images, lead to an increase of the rate of false positives. We found that, with different types of thresholds, even though we were able to detect the fainter stars, we got a greater number of false positives near the edges of the image. The application of filters to smooth the noise, coupled with an adaptive threshold might be able to reduce the false positives. Furthermore, as all tests have been performed with simulated data sets, we must still to validate the current routines over real data.

    \subsection{Impact of the star's rotation} \label{Sec:impact_rotation}

        As we have mentioned within this paper, we expect to find larger errors in the light curves for the orbiting stars, due to the continuous change in the pixels that are collecting light from the star. A portion of this degradation might come from the Flat Field correction applied to the images. Although expected to be a lower noise source, the FF correction being applied to the target stars, where it is assumed a single spectra type, will not match the background star.

        In order to quantify the loss in precision for those stars, when compared with the target star, we shall use Data Set C, which has 3 stars all with the same magnitude, as described in Appendix \ref{APP:datasets}. We had to decrease the magnitude of the closest star to Star 1 ($\ang{;;5}$,-6), to avoid having significant cross-contamination between the target and the background stars. Disregarding the previously set naming conventions, we shall refer to each star with the designation set in Fig.~\ref{init_pos_stars}, to maintain consistency between the different parts of the current Section. 

        \begin{figure*}
        \centering
                \includegraphics[width=13cm]{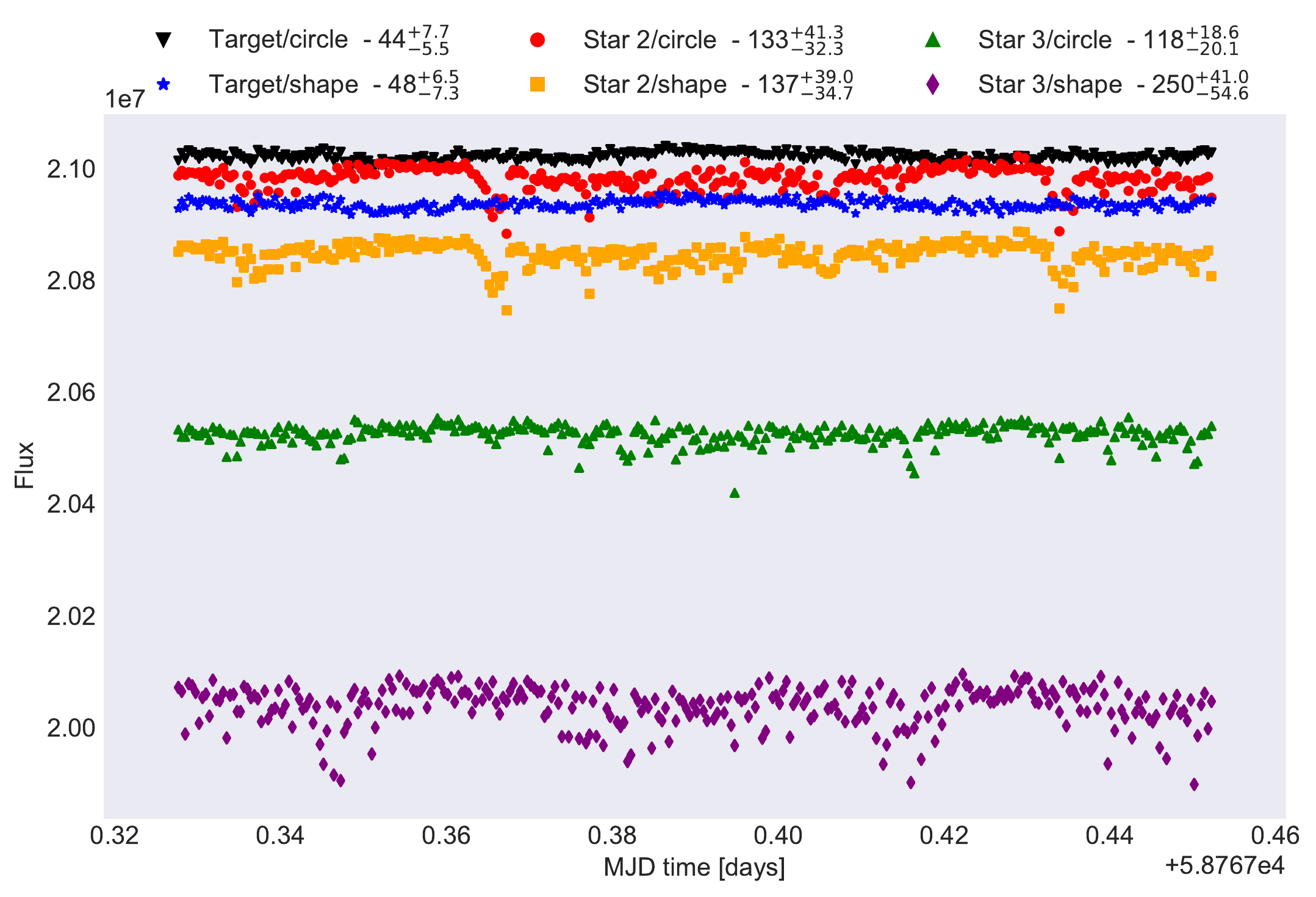}
                \caption{\label{C_targbg_compare}Light curves extracted, for Data Set C, without a background grid, a \textit{dynam} initial detection method and a \textit{dynam} star tracking method. In Table ~\ref{tab:targ_bg_comp} we can find the photometric precision and corresponding value of the error bars in its value.}
        \end{figure*}

        Furthermore, to guarantee consistency in the comparison between the target and the background stars, we shall refrain from using a background grid since it has an higher impact on the central star. Furthermore, we shall use the same initial detection method, \textit{dynam}, and star tracking method, \textit{dynam}, for all stars. In Section~\ref{Sec:compare_archi_methods} we have seen that this combination yielded the worst results for the target star, but the best ones for background ones.

        \renewcommand{\arraystretch}{1.3}
        \begin{table}
            \centering
            \caption{Comparison of the CDPP for the light curve extracted, with \textit{archi} and the DRP, from the target star, while using Data Set C.}
            \label{tab:targ_bg_comp}
            \begin{tabular}{lcc} 
            \hline 
            \multicolumn{2}{c}{Light curve }& Noise [ppm] \\
                \hline 

            \multirow{2}{*}{Target} & circle & $44^{+7.7}_{-5.5}$ \\
             & shape& $48^{+6.5}_{-7.3}$ \\
            \multirow{2}{*}{Star 2} & circle & $133^{+41.3}_{-32.3}$ \\
            & shape & $137^{+39.0}_{-34.7}$ \\

            \multirow{2}{*}{Star 3} &circle & $118^{+18.6}_{-20.1}$ \\
            & shape & $250^{+41.0}_{-54.6}$ \\

            \end{tabular}
        \end{table}

        From Fig.~\ref{C_targbg_compare} we can see, at a first sight, that Star 2 ($\ang{;;7}$,0) presents a periodic dip in the flux value, present in both masks. This periodic occurrence is due to the shape of the central star PSF that is, as previously discussed, crossed by the masks of the background stars. Some fluctuations can also be found for Star 3, but not as clearly. The light curve from Star 3 ($\ang{;;8}$,0) also presents slightly lower flux values, although that was expected due to the limitations imposed to the mask's size by the closeness to image's border. It's also noteworthy that, once again, we find that the \textit{shape} mask performs much worse than the \textit{circle} mask, for this star.

        If we now compare the photometric precision from the target star to the background ones, in Table~\ref{tab:targ_bg_comp}, we see that the former achieves a photometric precision  2 to 3 times better than the ones obtained for the background stars. The errors in the photometric precision also take smaller values for the light curve from the target star.

        Within the background stars we find similar noise values among the stars, although Star 3 ($\ang{;;8}$,0), with a circular mask, has lower error bars in the photometric precision. Furthermore, we can also see that the error bars suggest that the target star is less affected by the noise sources, which we can also verify with a visual comparison of the light curves.

        Finally, it's important to notice that even though the chosen combination of methods is not the optimal one for the target star, the photometric precision in it surpasses the one obtained for the background ones.

    \subsection{Contamination of the target by background stars}\label{APP:contam_targ_bg}

    In order to understand if the background stars will induce spurious signals on the target's star light curve we made use of Data Set D1 and D2, shown in Appendix \ref{APP:datasets}. The former, has a planet orbiting the background star, Star 1 ($\ang{;;5}$,-2), whilst the latter does not have it. If we subtract the light curve extracted from the target star of D2 from the one extracted on D1, we can see the portion of the astrophysical signal that is in fact introduced by nearby transits. If no contamination exists, then one would expect to find that both light curves are equal, without  exhibiting any differences between them.

        \begin{figure}
                \centering
                \includegraphics[width=9cm]{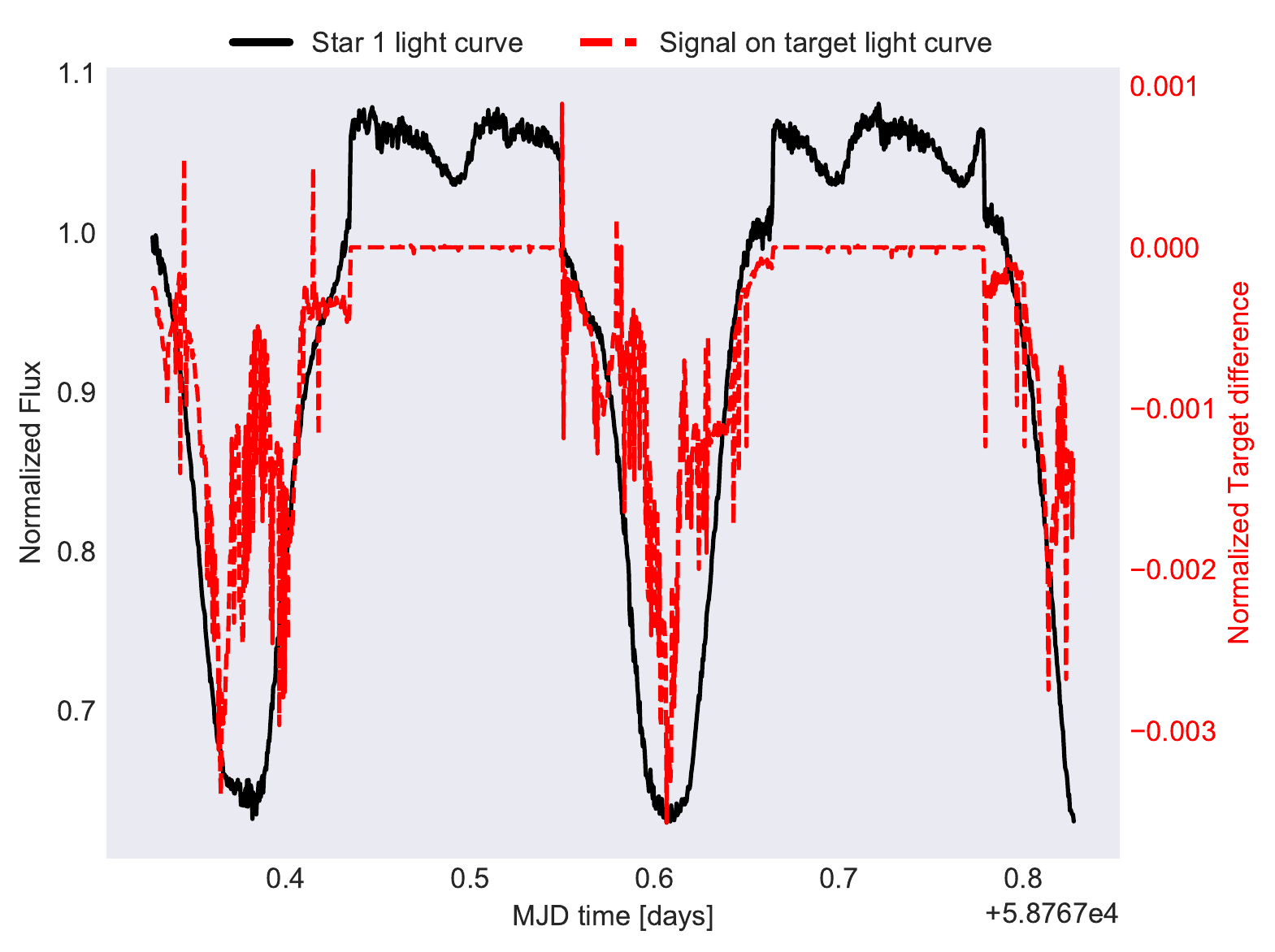}
            \caption{\label{Fig:signal_targ} Comparison of the signal found on the target star, with the transits of the background one. In black we have the normalized light curve from the background star, on Data Set D1, where we can clearly see a transiting event. In the red dashed line, we have the absolute difference between the normalized light curve from the central star, when there is a transit on the background star (D1) and when there isn't one (D2).}
        \end{figure}

    From Fig.~\ref{Fig:signal_targ} one can see that the residual signal is not zero but instead shows a behavior concordant with the transiting events from the background star. Furthermore, the fact that we can see effect of the background star's transit dip also serves as an indicator of its impact, even when there is no transit occurring.

    With this, we can conclude that it is possible to find spurious signals on the target's light curve, induced by other nearby stars. As \textit{archi} gives one the possibility of extracting those light curves, it allows for a manual analysis  to  validate if the astrophysical signal in question might derive or not from contamination of the background stars. At a later stage, those light curves could also be used to correct the contamination that each star applies on its neighbors, similarly to what was done for the K2 mission \citep{luger_update_2018}.

\section{Conclusions}

    Within this work we have presented \textit{archi}, an open-source python pipeline that is capable of extracting light curves from the background stars of the \textit{CHEOPS} mission. This new pipeline is capable of detecting and properly tracking the rotating stars, extracting light curves with a photometric precision 2 to 3 times worse than the one obtained for the target star. By resampling the original image we managed to create a virtual reduction in pixel size and thus achieved a reduction of the noise metric for the central star. Despite the good results for that star, the same was not verified for the background stars, for whom this grid would yield almost no gains.

    Through an empirical analysis we found that the benefits for the target star are also capped, i.e. using grids with a side with more than 1800 points does not reduce more the noise but, instead, the noise oscillates around an ``equilibrium'' value.  This equilibrium stage is met for the smallest possible grid, with 600 points, which was also deemed the grid's size that minimized the computational cost whilst providing light curves with that minimized the noise metric (CDPP). 

    We have also tested \textit{archi} in edge cases, i.e. saturated stars or very faint stars. Under cases where we find a saturated star, due to one of the background stars being brighter than the target, we can still detect fainter stars, albeit with an higher computational cost for the pipeline. We believe that the normalization routine, that represents the larger bottleneck, especially with larger background grids, can be improved by filtering the images to reduce noise and, consequently, the number of false positives. Furthermore, if we replace the binary threshold with an adaptive one coupled with smoothing filters, could be able to give better estimates of the masks.

    We have also found that in some cases the background stars could contaminate the target's light curve, imprinting variations that coincide with the transiting events of the background star. Since our pipeline is capable of extracting the light curves of background stars it allows to search them for possible causes of astrophysical signals. We postulate that a later stage it could be possible to use those light curves to remove noise from the target star, similarly to what was done for the K2 mission.

    Lastly, when comparing the light curves from the target star against the official Data Reduction Pipeline, which is still under development, we found that \textit{archi} is capable of outputting light curves with an equal or lower noise metric than the DRP, with a smaller 68\% confidence interval.

\section*{Acknowledgements}
We thank Mathias Beck, on behalf of CHEOPS SOC and David Futyan, for the work on the CHEOPSim tool. We also thank the anonymous reviewer for the comments that allowed us to improve the manuscript.

This work was supported by FCT - Fundação para a Ciência e a Tecnologia through national funds and by FEDER through COMPETE2020 - Programa Operacional Competitividade e Internacionalização by these grants: UID/FIS/04434/2019; UIDB/04434/2020; UIDP/04434/2020; PTDC/FIS-AST/32113/2017 \& POCI-01-0145-FEDER-032113; PTDC/FIS-AST/28953/2017 \& POCI-01-0145-FEDER-028953. S.S. acknowledge support from FCT through Investigador FCT contract nº IF/00028/2014/CP1215/CT0002 and in the form of an exploratory project with the same reference.
O.D.S.D acknowledge support from FCT through national funds in the form of a work contract with the reference DL 57/2016/CP1364/CT0004.




\bibliographystyle{mnras}
\bibliography{biblio} 



\appendix

\section{Data Sets in use}\label{APP:datasets}

    For this work we used 7 different data sets, for three distinct use cases. The first, Data Set A, was used to compare \textit{archi}'s methods amongst themselves and with the DRP. The following three, Data Set B1,B2 and B3, were used to see how the pipeline behaves for different exposure times.  Data Set C was created to study the difference between the precision obtained for the target star and for the background ones, when the magnitude of all visible stars is equal. Finally, Data Set D1 and D2 were used to show that background stars could contaminate the light curve from the target star.

\begin{figure}
            \centering
            \includegraphics[width=8cm]{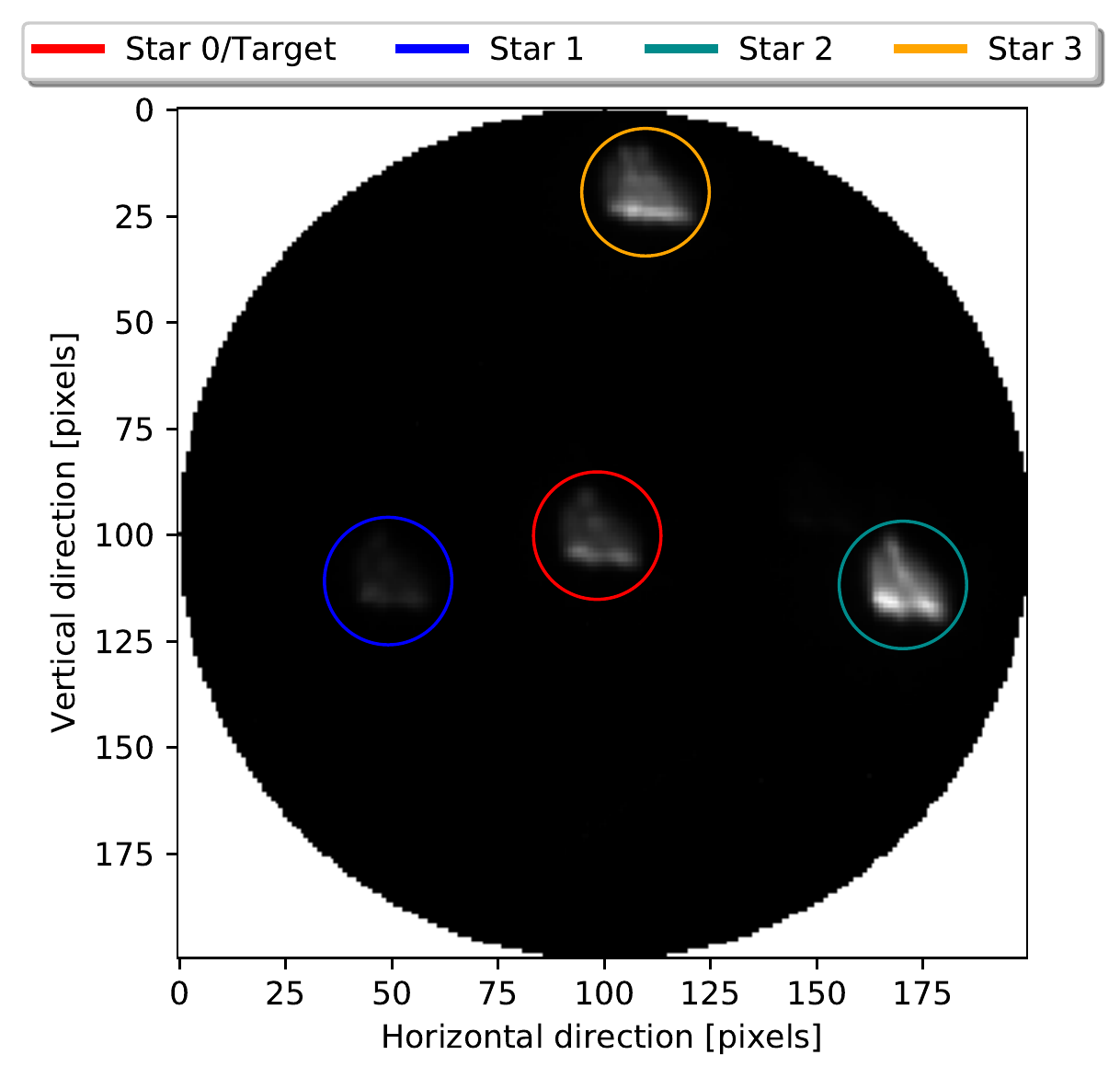}
        \caption{\label{init_pos_stars} Initial position of the stars, for all Data Sets.}
    \end{figure}

    We maintained the separation between the background stars constant, as seen in Fig.~\ref{init_pos_stars}, albeit with different magnitudes, as one can see in Table~\ref{tab:stars_mags}, depending on the use case of the respective Data Set. Data Set B1, B2 and B3 all have equal magnitudes of the background stars, with the only difference being on the magnitude of the target star and, consequently, the exposure time of the images. In Data Set C we made the magnitude of Star 1 such that it was too faint to be visible in the images. Lastly, for Data Set D1 and D2 we have also made Star 2 and Star 3 fainter, so that they would not appear on the images and, in Data Set D1 we introduced a transiting event on Star 1.

    \begin{figure}
            \centering
            \includegraphics[width=8cm]{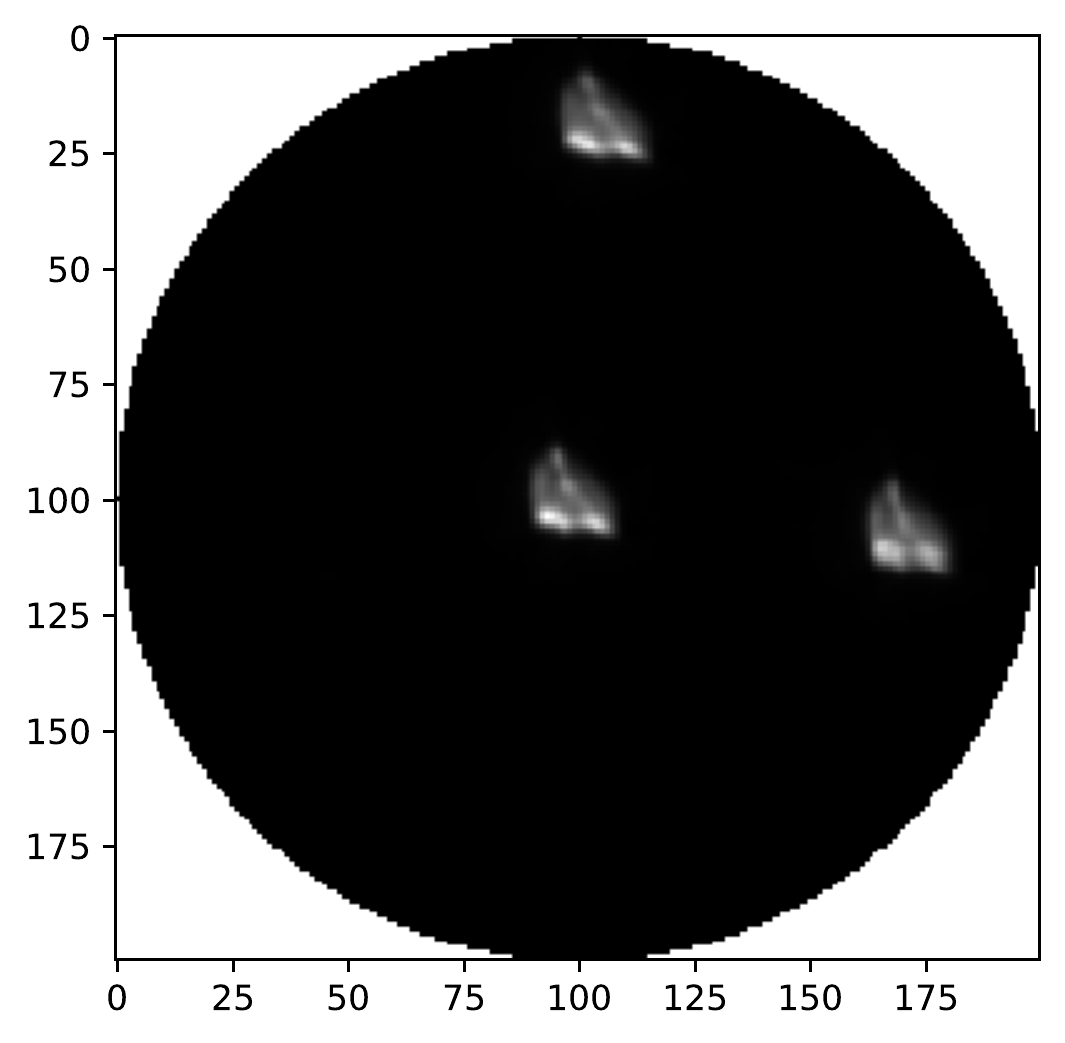}
        \caption{\label{stars_C}Stars present on Data Set C.}
    \end{figure}   

    \begin{figure}
            \centering
            \includegraphics[width=8cm]{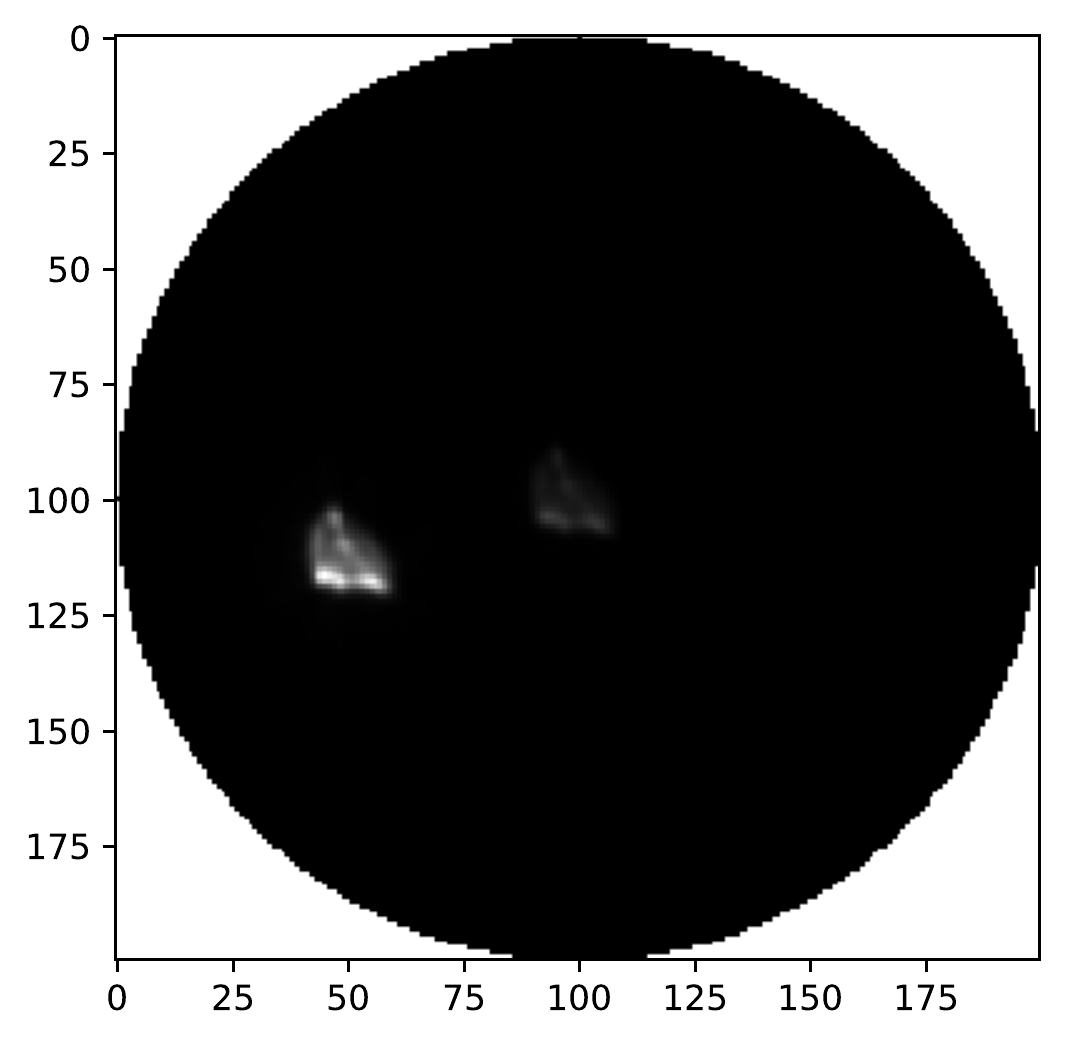}
        \caption{\label{stars_D} Stars present on Data Set D.}
    \end{figure}

    \begin{table*}
    \centering
    \caption{Mapping between the stars, seen in Fig.~\ref{init_pos_stars}, their magnitudes in each data set and the separation from the target star, in arcseconds. It's important to keep in mind that Data Set C has Star 1 set in such a way that it does not appear in the images. It's important to keep in mind that, in this Data Set, Star 1 will not visible in the images and, consequently not detected by \textit{archi}.}
    \label{tab:stars_mags}
    \begin{tabular}{lccccc} 
        \hline
         Data Set& Star 0/Target [mag]& Star 1 [mag]& Star 2 [mag]& Star 3 [mag]& Exposure time [s] \\
        \hline
        A & 12  &13 & 11 & 11.5 &  60\\  
        B1 & 7 & 13 & 11 & 9 & 22\\    
        B2 & 9 & 13 & 11 & 9 & 36 \\     
        B3 & 12 &13 & 11 & 9 & 60\\  
        C & 9 & 15 & 9& 9 & 36\\     
        D1/D2 & 9 & 7 & --& -- & 36\\  
        \hline
        Distance to centre[arcsec] &0& 5.004 & 7.300 & 8.155 & --\\
        Transiting events & No& In D1& No & No& -- \\
        \hline
    \end{tabular}
\end{table*}


\bsp	
\label{lastpage}
\end{document}